\documentclass[aps,pre,twocolumn,superscriptaddress]{revtex4-1}
\usepackage{amsmath}
\usepackage{amssymb}
\usepackage{color}
\usepackage{graphicx}
\usepackage{epstopdf}

\begin{document}

\title{Light-Induced Manipulation of Passive and Active Microparticles}

\author{Pooja Arya}
\affiliation{Institute of Physics and Astronomy, University of Potsdam, 14476 Potsdam, Germany}

\author{Maren Umlandt}
\affiliation{Institute of Physics and Astronomy, University of Potsdam, 14476 Potsdam, Germany}

\author{Joachim Jelken}
\affiliation{Institute of Physics and Astronomy, University of Potsdam, 14476 Potsdam, Germany}

\author{David Feldmann}
\affiliation{Tel Aviv University,  Israel}

\author{Nino Lomadze}
\affiliation{Institute of Physics and Astronomy, University of Potsdam, 14476 Potsdam, Germany}

\author{Evgeny S. Asmolov}
\affiliation{Frumkin Institute of Physical Chemistry and Electrochemistry, Russian Academy of Sciences, 31 Leninsky Prospect, 119071 Moscow, Russia}

\author{Olga I. Vinogradova}
\email[Corresponding author: ]{oivinograd@yahoo.com}
\affiliation{Frumkin Institute of Physical Chemistry and Electrochemistry, Russian Academy of Sciences, 31 Leninsky Prospect, 119071 Moscow, Russia}
\affiliation{DWI - Leibniz Institute for Interactive Materials,  Forckenbeckstr. 50, 52056 Aachen, Germany}

\author{Svetlana Santer}
\email[Corresponding author: ]{santer@uni-potsdam.de}
\affiliation{Institute of Physics and Astronomy, University of Potsdam, 14476 Potsdam, Germany}

\date{\today }

\begin{abstract}
We consider sedimented at a solid wall particles that are immersed in water containing small additives of photosensitive ionic surfactants.  It is shown that illumination with an appropriate wavelength, a
beam intensity profile, shape and size could lead to a variety of dynamic, both unsteady and steady-state,
configurations of particles.  These dynamic,  well-controlled and switchable particle patterns at the wall are due to an emerging diffusio-osmotic flow that takes its origin in the adjacent to the wall electrostatic
diffuse layer, where the concentration gradients of surfactant are induced by
light. The conventional nonporous particles are passive and can move only with already generated flow. However,    porous colloids actively participate themselves in the flow generation mechanism at the wall, which also sets their interactions that can be very long ranged.
This light-induced diffusio-osmosis opens novel avenues
to manipulate colloidal particles and assemble them to various patterns. We show in particular how to create and split optically the confined regions of particles of tunable size and shape, where well-controlled flow-induced forces on the colloids could result in  their crystalline packing, formation of  dilute lattices of well-separated particles, and other states.

\end{abstract}

\maketitle

\affiliation{Frumkin Institute of Physical Chemistry and Electrochemistry,
Russian Academy of Sciences, 31 Leninsky Prospect, 119071 Moscow, Russia}
\affiliation{DWI - Leibniz Institute for Interactive Materials,
Forckenbeckstr. 50, 52056 Aachen, Germany}

\affiliation{Frumkin Institute of Physical Chemistry and Electrochemistry,
Russian Academy of Sciences, 31 Leninsky Prospect, 119071 Moscow, Russia}


\section{Introduction}

The ability to dynamically manipulate colloidal particles at interfaces is essential prerequisites for offering new and revolutionary solutions for developing strategies of self-repairing complexes and integrated structures, the creation of new materials such as hybrid bio-electro-mechanical systems, building adaptive sensors, optoelectronic, optical devices, active structural color materials~\cite{Zhang.H:2020,Wang.W:2019,Bian.F:2020,Wong.T:2011,Rogers.J:2010,Fennimore.A:2003}.   To generate colloidal ensemble in temporary steady-states rather than creating static architectures  different strategies has been developed~\cite{Liljestrom.V:2019}.  For instance employing external electric or magnetic fields~\cite{Fu.Z:2016,Crassous.J:2014,Sherman.Z:2018},
electrostatic~\cite{Li.G:2006,Demirors.A:2018}, optothermal~\cite{Lin.L:2018}, optofluidic~\cite{caciagli.a:2020},
  and hydrodynamic~\cite{Pedreroa.F:2018} interactions  it is possible to trigger the assembly of colloids into adjustable functional pattern~\cite{Chiou.P:2005,Dienerowitz.M:2008,Yang.A:2009,Rodriguez.N:2004,Lee.C:2001,Ruan.G:2010}.

  Extensive efforts have gone into investigating particles at the gas-liquid and liquid-liquid interface~\cite{lv.c:2018,caciagli.a:2020,dominiges.a:2016,peter.t:2020}. The body of theoretical and experimental work
studying particles at the solid surface is much less than that for gas-liquid interface, although
there is a growing literature in this area. For instance, it has been shown that a self-generated solvent flow  at the solid-liquid interface can be used to establish long-range attractions on the colloidal scale leading to the aggregation \cite{niu.r:2017}. Electro-osmotic flow near the wall has been reported to drive sedimented particles~\cite{niu.r:2017b}. Different mechanisms of the colloidal motion can be induced by employing the
so-called diffusioosmosis (DO), where the flow tangent to the wall is driven by the gradient of
the solute concentration (or the osmotic pressure gradient)~\cite{prieve.dc:1984,Marbach.S:2017}.
It is
traditionally considered that the DO flow takes its origin in the adjacent
to the wall thin surface layer, such as, for example, the diffuse layer of
counter-ions. If the settled at the wall particles dissolve into ions, this produces local ion
gradients driving microflows and particle movement along the charged
solid surface~\cite{mcdermott2012}. The DO flow can also be induced by light allowing to manipulate inert (i.e. nondissolvable) particles~\cite{feldmann2016}, and it has recently been reported that some type of particles, such as porous, can even participate in its generation mechanism, acting as micropumps and thus representing a novel type of active colloids~\cite{feldmann.d:2020}.

Previous studies of light-driven DO, usually referred to as LDDO, have addressed the motion and aggregation of either passive or active particles under illumination of focussed or uniform light \cite{feldmann2016,feldmann.d:2020}.
Here we show how to induce changes in dynamic pattern of colloidal ensemble of nonporous (passive) and porous (active) particles, as well as temporary steady states, by using the irradiation with light of combined wavelength and simple adjustment of its parameters, such as wavelengths, intensity and its distribution, and the size of the laser beam. Since porous particles can generate the LDDO flow near the wall themselves leading to their repulsion or attraction, and this is easily controlled by irradiation parameters, the variety of surface patterns becomes extremely rich. Our work opens up many intriguing avenues, such as manipulating  ensembles of different particles with fine-tuning of both their interaction forces and confining  areas.

Our paper is arranged as follows. In Sec.~\ref{sec:concept} a general concept of light-induced diffusio-osmosis is introduced.
Some general considerations concerning DO velocities of a liquid induced by the laser spot that drives passive and active particles is presented.
The description of the experimental system, materials and methods used is given in Sec.~\ref{Sec:experimental}. Sec.~\ref{Sec:results} describes the experimental results. Here we explore and contrast the light-induced aggregation properties of active and passive particles, and show that by tuning the irradiation with a combined wavelength one can remotely control their  assembly and, as a result, generate complex dynamic patterns of particles at the solid surface.  We conclude in Sec.~\ref{Sec:conclusion}.

\section{General concept of light-driven diffusio-osmosis}\label{sec:concept}

We first present the theoretical basis and physical ideas underlying the
approach, which we have developed before and slightly extend here to manipulate
microparticles, sedimented under gravity to the bottom wall, but not really
stick, as known from numerous hydrodynamic and electrokinetic experiments \cite%
{Williams1992,asmolov2015,dubov.al:2017}. Therefore, they can move along the wall
when a lateral force is applied. In our experiment the particles are driven by the fluid DO flow generated by the photosensitive surfactant concentration gradient near a bottom wall. The velocity of this flow, $v_{DO}$, is generally not equal to that of particles, $v_p$. Below
we estimate these theoretically, by focusing on two specific limiting situations. Namely, of monodisperse suspensions of passive impenetrable particles illuminated by  a focused beam and that of permeable porous particles subject to a homogeneous unfocused illumination. The understanding of basic principles underlying the migration of particles in these two simple (ideal) system configurations is  important to design of  experiments we report below (Sec.~\ref{Sec:results}), where more complex laser beams are employed to manipulate particles to provide their assembly to various patterns.

\subsection{Passive particles. }\label{sec:theory_passive}

\begin{figure}[tbp]
\begin{center}
\includegraphics[width=1.0\columnwidth]{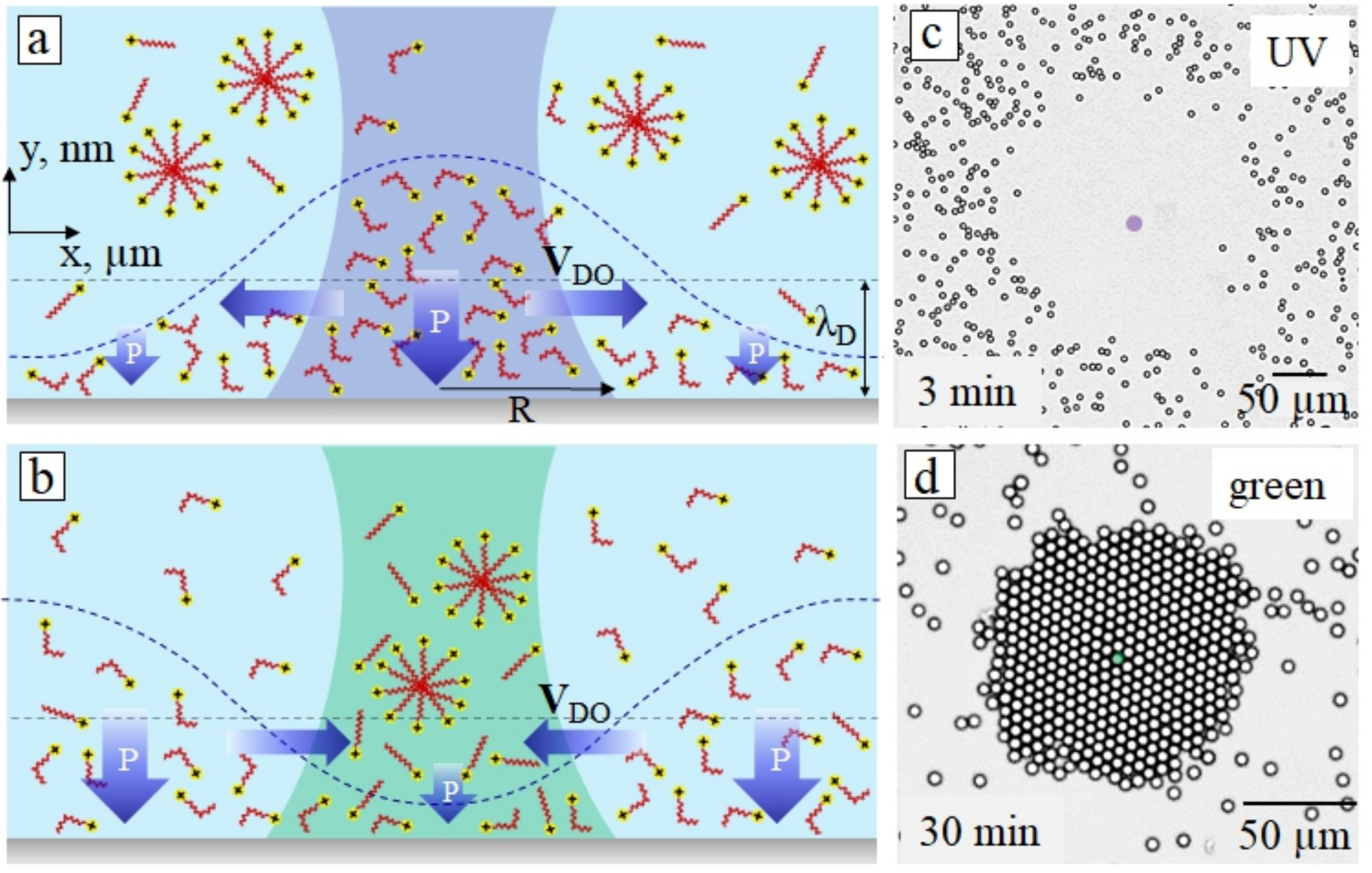}
\end{center}
\par
\vspace{-0.4cm}
\caption{(a)  Schematic of the  DO flow induced by  focused irradiation with UV light. The flow is directed outwards of the laser beam. (b) The same for a flow inwards that is generated with green light. The dashed line indicates distribution of \emph{cis}-isomer concentration. (c,
d) Examples of the optical micrographs showing removal (c) and gathering (d)
of the passive particles (adapted from Ref.~\cite{feldmann2016}). }
\label{Fig:sketch}
\end{figure}

The first typical case we address is the radially directed DO flow. Such a situation happens when an electrolyte
solution containing photosensitive (azobenzene) ionic surfactant is
irradiated with the focused light as shown in Fig.~\ref{Fig:sketch}. The particle radius $a$ in such experiments is typically of several $\mu$m, and the laser spot radius $R$ is of the order of ten $\mu$m. Since only a portion of the wall is irradiated, by choosing an appropriate wavelength one
can generate \emph{trans}-isomers inside the focussed beam and \emph{cis} - outside of it,
and vice versa~\cite{feldmann2016}. For example, focused UV light (converting \emph{trans}- to \emph{cis}-) induces the flow out of the irradiated area as seen in Fig.~\ref{Fig:sketch}(a, c). By contrast,  when the solution of \emph{cis}-isomers is irradiated with focused green
light (converting fast \emph{cis}- to \emph{trans}-) (see Fig.~\ref{Fig:sketch}(b, d)), the flow is directed
towards the laser spot. Note that above a certain critical (bulk) micelle concentration, abbreviated as $cmc$, the majority of \emph{trans}-isomers finds itself aggregated in micelles.

Near the bottom wall an electrostatic diffuse layer, i.e. a cloud of
counter-ions balancing its surface charge, is formed. A measure of the thickness of the diffuse layer is the Debye screening length of the solution, $\lambda _{D}$, and the excess of mobile (thermal) ions in the diffuse layer compared to the bulk can be denoted as $\Gamma(x)$, where the $x$%
-axis is defined along the wall. The value of $\Gamma(x)$ can be related to the surface potential of the  wall, $\phi _{s}$, as~\cite{feldmann2016}

\begin{equation}
\Gamma(x)\simeq c_{0}\lambda _{D}\left( \dfrac{e\phi _{s}}{k_{B}T}\right)^{2},  \label{Eq:Gamma_ed3}
\end{equation}%
where  $c_0$ is the bulk concentration of surfactants, $e$ is the elementary (positive)
charge, $k_{B}$ is the Boltzmann constant, $T$ is the temperature. Naturally, in the absence of irradiation $%
\Gamma$ does not depend on $x$, and the system is at equilibrium. However, when it is illuminated, the lateral gradient of the excess (mobile) surfactant ions in the diffuse layer
$\partial \Gamma /\partial x$ induces the liquid flow of the velocity~\cite{feldmann2016}

\begin{equation}
v_{DO}\simeq \dfrac{\lambda _{D}}{\eta }k_{B}T\left[ \dfrac{\partial \Gamma%
} {\partial c_{t}}\nabla c_{t}(x)+\dfrac{\partial \Gamma}{\partial c_{c}}%
\nabla c_{c}(x)\right] ,  \label{velocity_DO}
\end{equation}%
with the index $\{t,c\}$ standing for  \emph{trans}- and  \emph{cis}-isomers, and derivatives $\partial \Gamma/\partial c_{t, c}$ characterizing the enrichment a diffuse layer by \emph{trans}- and \emph{cis}- molecules.

Below $cmc$ the surfactants exist in a form of isolated molecules, so that $c_{c}=c_{0}-c_{t}$, and Eq.(\ref%
{velocity_DO}) can then be reduced to

\begin{equation}
v_{DO}\simeq \dfrac{\lambda _{D}}{\eta L} c_0 k_{B}T \left( \dfrac{\partial \Gamma
}{\partial c_{c}}-\dfrac{\partial \Gamma}{\partial c_{t}}\right), \label{velocity_DO2}
\end{equation}%
where we assumed that $\nabla
c_{c}\simeq c_0/L$ with the length scale of inhomogeneous
concentration field $L$. It is evident that $L$ is roughly equal to the radius of the laser spot $R$ for a configuration shown in Fig.~\ref{Fig:sketch}, but it can be smaller for a more complex irradiation when finite $\nabla
c_{c}$ is generated only in some portion of the beam. We will return to this point in Sec.~\ref{sec4a}. Note that the velocity is also controlled by the difference  $(\partial \Gamma/\partial c_{c}-\partial \Gamma/\partial c_{t})$ that can be evaluated by measuring the surfactant adsorption isotherm and the wall potential. Here we simply assume that it is of the order of $\Gamma/c_0$.  Overall, we conclude that the magnitude of $v_{DO}$ should grow linearly with $c_0$, and that the direction of flow is generally defined by the enrichment of the diffuse layer by \emph{trans}- or \emph{cis}-isomers. Another conclusion from Eq.~\eqref{velocity_DO2} is that $v_{DO}$ reduces with added salt. All this was indeed observed in prior experiments. Finally, we note that $P = c_0 k_{B} T$ is the osmotic pressure of a surfactant solution, so that everything can be equally discussed in terms of variations in osmotic pressure. In Fig.~\ref{Fig:sketch} and some further figures we have included the arrows of different size that intend to indicate the value of $P$ (but, importantly, not a direction of the pressure force) in different part of the diffuse layer.

When $c_0$ reaches $cmc$, a fraction of \emph{trans}-isomers does remain in a non-aggregated form. Their concentration is constant ($\nabla
c_{t}\simeq 0$) and equal the $cmc$ so as to maintain the micelle/\emph{trans}-isomer equilibrium. Eq.(\ref%
{velocity_DO}) can be then transformed to

\begin{equation}
v_{DO}\simeq \dfrac{\lambda _{D}}{\eta }k_{B} T\dfrac{\partial \Gamma}{%
\partial c_{c}}\nabla c_{c}(x).  \label{velocity_DO3}
\end{equation}%
Then $v_{DO}$ can be estimated using Eq.\eqref{Eq:Gamma_ed3}
\begin{equation}
v_{DO}\simeq \dfrac{\lambda _{D}}{\eta }\dfrac{k_{B}T\Gamma}{L}\simeq \dfrac{%
 c_0 k_B T \lambda _{D}^{2}}{\eta L}\left( \dfrac{e\phi _{s}}{k_{B}T}\right)
^{2}.  \label{velocity_DO4}
\end{equation}
Note that since surfactants are ionic, at much higher than $cmc$ concentrations they can simultaneously reduce $\lambda _{D}$ and increase $\eta$, and, consequently, decrease $v_{DO}$, compared to predictions of Eq.\eqref{velocity_DO4}. Some experimental measurements~\cite{feldmann2016} lend some support of this. Therefore, to maximize $v_{DO}$ it would be reasonable  to keep $c_0 \simeq cmc$ and to avoid high concentration of surfactants. Another advice would be to use pure water to provide largest possible $\lambda_D$.

Finally, we stress that since the laser spot radius $R$ is normally of the order of ten $\mu$m, the fluid velocity could be of the order of
ten $\mu $m/s. This long-range flow appears as homogeneous on the scale of small passive (i.e. smooth impermeable) particles
of size $a$. Such a flow captures them, and particles move with the velocity comparable to that of the DO flow, $%
v_{p} \simeq v_{DO}$.

The previous consideration is, of course, grossly simplified, but it provides us with some guidance on a control of DO velocities.  We also recall that our model assumes ideal, impenetrable bottom wall, typical for most ordinary applications. In this case the (dynamic) $\zeta$-potential, which, in fact,  controls the induced velocity, is equal to $\phi_s$. Therefore, one can increase the velocity by using coatings that are permeable to water and ions, such
as porous or rough ones, for which  $\zeta$-potential is normally augmented compared to the surface electrostatic potential~\cite{vinogradova.oi:2020}, which should lead to an increase in $v_{DO}$ given by  Eq.\eqref{velocity_DO4}.

\subsection{Active particles. }\label{sec:theory_active}

\begin{figure}[tbp]
\begin{center}
\includegraphics[width=1.0\columnwidth]{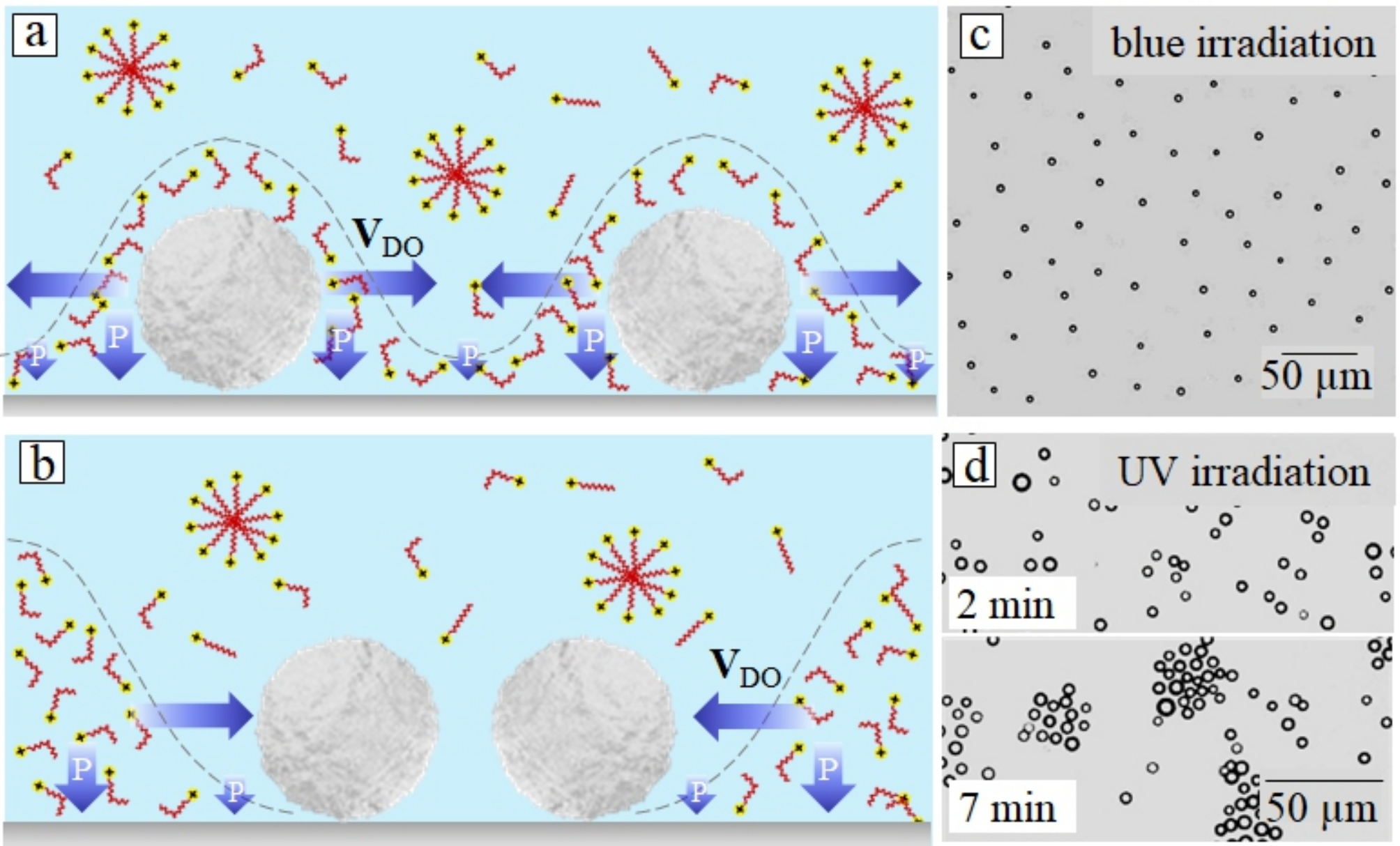}
\end{center}
\par
\vspace{-0.4cm}
\caption{(a) Schematic of a DO repulsion of two active porous particles under uniform irradiation with blue light. (b) The same of a DO attraction under irradiation with UV
light. (c, d) Examples of the optical micrographs showing a lattice of porous particles separated by distances of several tens of $\mu$m (c) and aggregation of porous particles due to their DO attraction (d) (adapted from Ref.~\cite{feldmann.d:2020}). }
\label{Fig:sketch2}
\end{figure}

An unfocused uniform illumination cannot, of course, drive the passive impermeable particles described in Sec.~\ref{sec:theory_passive} since in this situation $\partial \Gamma /\partial x = 0$. However, a light-induced DO flow can be generated when sedimented particles are porous~\cite{feldmann.d:2020}. At equilibrium such particles absorb the ionic species~\cite{silkina.ef:2020,rumyantsev.a:2014,schimka.s:2017a,schimka.s:2017b}, but  irradiation with blue light induces a rapid escape of \emph{cis}-isomers out the pores. Consequently, each particle becomes a source of a laterally inhomogeneous excess of \emph{cis}-isomers that leads to a concentration inhomogeneity along the wall. This in turn induces the (local) flow along the wall, which is directed away from the particle. In other words, the porous particle acts as a micropump, so that it can be seen as active. An illuminated isolated particle cannot, of course, translate due to a symmetry of emerging flow. However, a region of enhanced pressure near a bottom pole of the porous particle with excess of \emph{cis}-isomers should inevitably induce a buoyant force (Fig.~\ref{Fig:sketch2}(a)). Consequently, such particles effectively become  lighter than passive ones, and, therefore, have smaller reaction force at support. Besides, porous particles are slippery~\cite{beavers.gs:1967,Feldmann.D:2019}. All these should reduce their contact friction on the bottom wall. One can, therefore, speculate that when the flow is induced by the focused beam (as in Sec.\ref{sec:theory_passive}), the porous particles would translate faster than nonporous ones.

In the case of an ensemble of particles at the wall they repel (Fig.~\ref{Fig:sketch2}(a)). Such a DO repulsion of active porous particles can be used for a remote control of their two-dimensional assemblies at the solid wall, and, in particular, simply by using different illumination wavelengths (eg. UV light) one can reversibly switch the state of a suspension of porous particles from a periodic lattice of (separated by distances on the order of tens of micrometers) particles to densely packed  surface aggregates (Fig.~\ref{Fig:sketch2}(b, c))\cite{Arya.P:2020b}. Note that a similar mechanism of a flow generation near active particles has been reported for calcium carbonate particles~\cite{mcdermott2012}. An important difference in our case is
that the magnitude and direction of the flow is fully controlled by the
intensity and wavelength of irradiation, and that the whole process is reversible.

The phenomenon is identical to described in Sec.~\ref{sec:theory_passive}, but now the length scale of the local DO flow is the particle size $a,$ so that the
velocity becomes larger than that for the flow induced by the focused light:
\begin{equation}
v_{DO}\simeq \dfrac{c_{0} k_B T \lambda _{D}^{2}}{\eta a}\left( \dfrac{e\phi _{s}%
}{k_{B}T}\right) ^{2}.  \label{u_active}
\end{equation}%
The axisymmetric flow around a single particle can not, of course, induces its
motion, but can drive the motion of neighbouring passive particles of a smaller size. We stress that in this case the velocities of particles could be significantly augmented compared to discussed in Sec.~\ref{sec:theory_passive}, where the DO flow was induced by focused laser beam~\cite{feldmann.d:2020}.

It is naturally to assume that the fluid velocity near the particle inversely decays with the distance. The superposition of such flows induced by neighbouring active
porous particle leads to their long-range DO
interactions, and the relative velocity of interacting particles is given by
\begin{equation}
v_{p}\simeq v_{DO} \dfrac{a}{d} \simeq \dfrac{c_{0} k_B T\lambda _{D}^{2}}{\eta d}\left(
\dfrac{e\phi _{s}}{k_{B}T}\right) ^{2},  \label{velocity_DO5}
\end{equation}%
where $d$ is the characteristic interparticle distance.
The mutual repulsions of many active particles result in a formation of a stationary periodic lattice of particles
that are separated by distances on the order of tens of micrometers (see Fig.~\ref{Fig:sketch2}(c)).
 Forming such a lattice particles  exhibit some hydrodynamic fluctuations resembling the Brownian motion.

\section{Experimental}\label{Sec:experimental}

\subsection{Materials}

\begin{figure}[tbp]
\begin{center}
\vspace{0.4cm} \includegraphics[width=1.0\columnwidth]{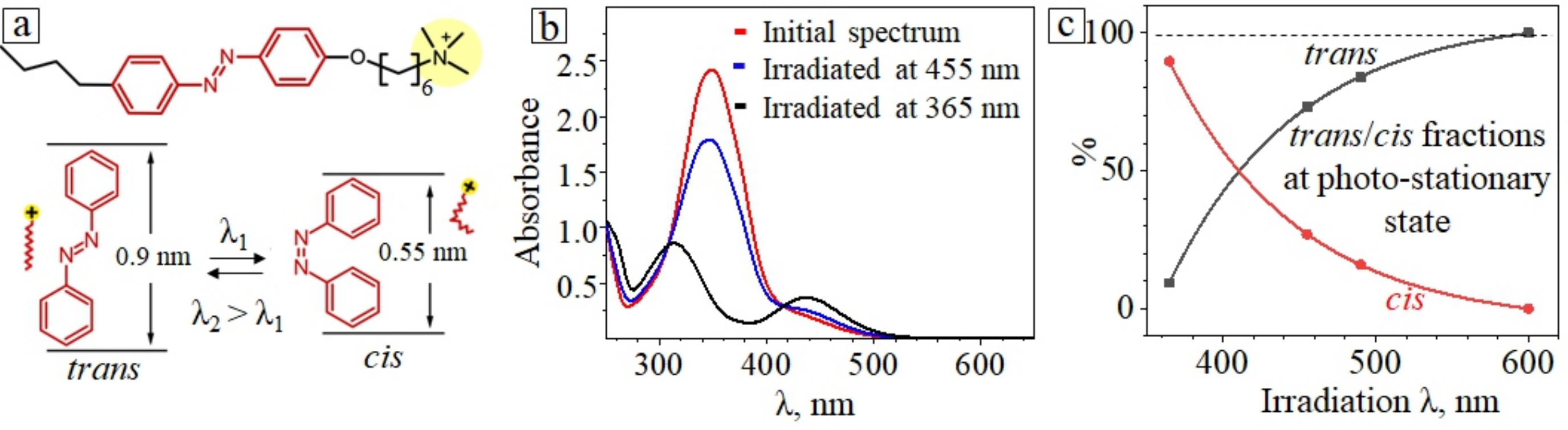}
\end{center}
\par
\vspace{-0.4cm}
\caption{(a) The chemical structure of the
photo-sensitive surfactant and two azobenzene isomers. (b) UV-Vis absorption spectra of a solution of $c_0$ = 1 mM irradiated with UV (black curve), blue (blue curve) light and in dark (red curve). (c) The corresponding amount of \emph{trans}-(black curve) and \emph{cis}-isomers (red curve) at a photo-stationary state as a function of irradiation wavelength. }
\label{Fig:surfactant}
\end{figure}

Azobenzene containing cationic surfactant consists of charged head (trimethyl-ammonium bromide) and hydrophobic tail in which azobenzene unit is incorporated~\cite{Dumont.D:2002} (see Fig.~\ref{Fig:surfactant}(a)).  The azobenzene undergoes photo-isomerization from the \emph{trans}- to the \emph{cis}-state under irradiation with UV light ($\lambda$ = 365 nm). The photo-stationary state with 90\% of \emph{cis}- isomers is achieved within several minutes (intensity dependent) of irradiation. UV-Vis absorption spectra recorded in dark (\emph{trans}- state) \cite{Arya.P:2020a} and during UV illumination at the photo-stationary state (\emph{cis}- isomers) are presented in Fig.~\ref{Fig:surfactant}(b). Under blue light, the \emph{trans}/\emph{cis} ratio at the photo-stationary state is 67\%/33\%. Under illumination with longer wavelengths (green light), the photo-isomerization from the \emph{cis}- to the \emph{trans}- state takes place within seconds, while thermal back relaxation in the dark takes more than 48 hours \cite{Arya.P:2020a}. The red light ($\lambda$ = 625 nm) illumination does not affect photo-isomerization of the surfactant.

The aqueous dispersion of nonporous silica particles of $a=2.5$ $\mu$m (purchased from microparticle GmbH, Germany)   or mesoporous silica particles of the same radius (pore diameter: 6 nm and BET value = 850 m$^2$/g , purchased from micromod, Germany) is mixed with surfactant aqueous solution of $c_0 = 1$ mM, which is twice above $cmc$. The particle concentration is adjusted to 0.1 mg/ml. The dispersion is kept for equilibration at least for 1 hour and later introduced to glass sealed chamber of height ca. 1 mm and sample volume of 40 $\mu$l. All samples are kept in the dark or in red light to prevent undesired photo-isomerization. We stress that no salt is added to maximize the velocities as discussed in Sec.~\ref{sec:concept}.

\subsection{Methods}

An inverted microscope (Olympus IX73) equipped with UV (M365L2-C1, Thorlab Gmbh), blue M455L3, Thorlab Gmbh) and red (M625L1- C1, Thorlabs Gmbh) light source is used for optical measurements.  Additionally, three lasers ($\lambda$ = 532 nm and 488 nm, Cobolt, Sweden; $\lambda$ = 375 nm, Coherent.inc, USA) are incorporated in microscope where all the beams are focused through the objective to the sample. The illuminated intensity of light is measured using optical power meter PM100D with sensor S170C (Thorlabs Gmbh, Germany). Micrographs are acquired with a Hamamatsu Orca-Flash 4.0 LT (C11440) at a rate of 1 frame per sec. The setup is kept in the dark to prevent the uncontrolled isomerization. When required, red light (M625L1- C1, Thorlabs Gmbh) is used for imaging in dark as it does not affect the photo-isomerization, i.e. the azobenzene molecules do not change their isomerization state. The intensity of irradiation for all light source is kept constant during the experiments.

The kinetics of photo-isomerization is measured using UV-Vis spectroscopy (Cary 5000 UV-Vis-NIR spectrophotometer, Agilent Technologies, USA) \cite{Arya.P:2020a}.


Image processing and analysis of image data is performed by Image J plugins (Mosaic Single Particle Tracking) using the tracking algorithm described by~\citet{Sbalzarini.I:2005}. 

\section{Results and discussion}\label{Sec:results}

\subsection{Choice of experimental parameters}
\label{sec4a}

\begin{figure}[tbp]
\begin{center}
\vspace{0.4cm} \includegraphics[width=1.0\columnwidth]{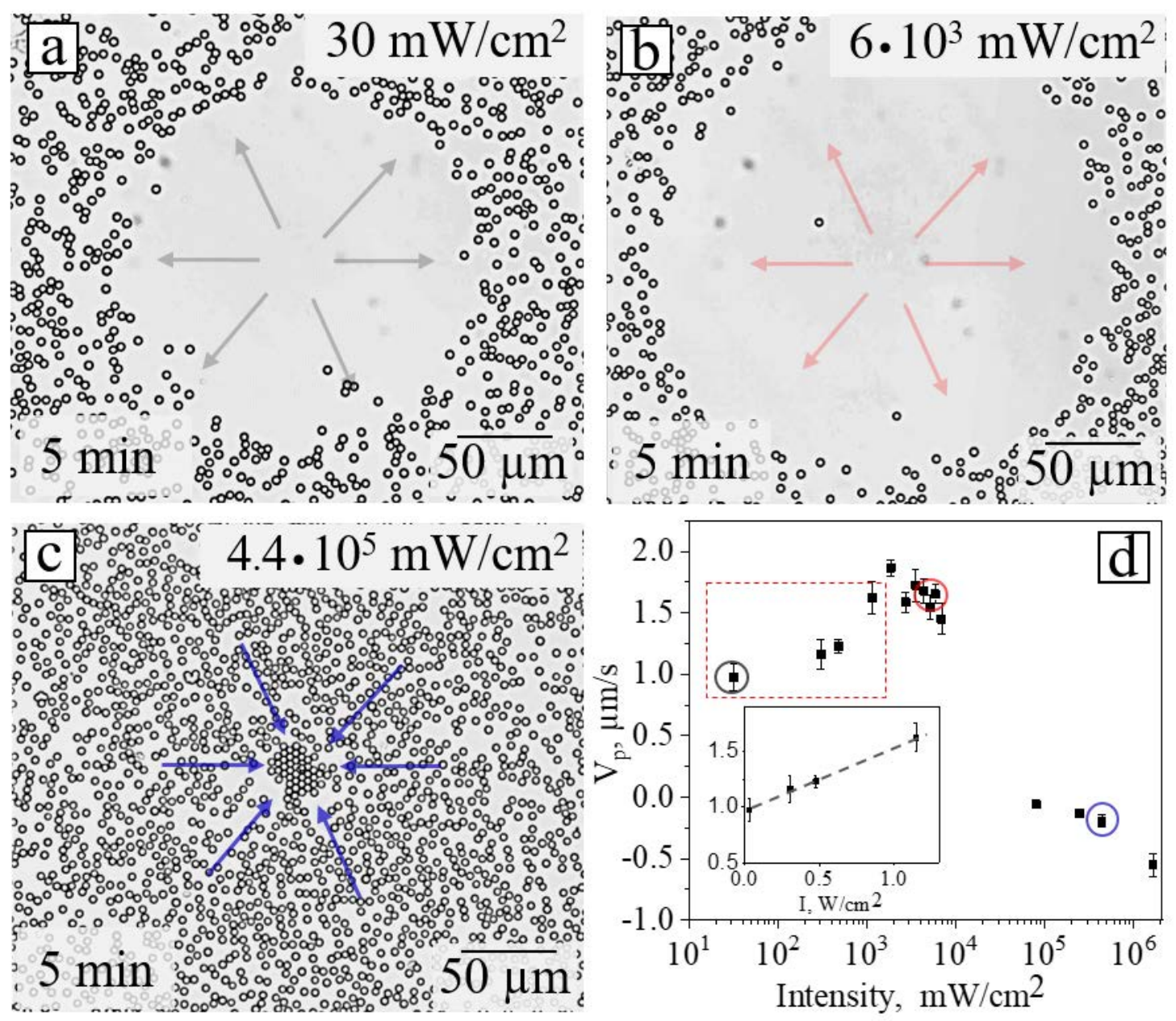}
\end{center}
\par
\vspace{-0.4cm}
\caption{(a,b,c) Optical micrographs of passive silica particles after 5 min irradiation with focused UV light ($\lambda$ = 375 nm) at different intensities: (a) 30 mW/cm$^2$, (b) $6 \times 10^3$ mW/cm$^2$, (c) $4.4 \times 10^5$ mW/cm$^2$. Arrows show the direction of LDDO flow. (d) A lin-log plot of the average particle velocity, $v_p$, in the ring near the edge of the cleaned circle as a function of light intensity. The circles (grey, red, blue) mark the intensities used in (a,b,c). The dashed rectangle constrains the range of intensities for further measurements. The solid rectangle represents an inset with a lin-lin plot of data confined in this dashed rectangle. The dashed line is a guide to the eye.}
\label{Fig:intensity}
\end{figure}

We first ascertain the range of intensities leading to the LDDO flow. Note that since the latter is the consequence of  absorption of the optical energy by dissolved surfactant molecules, at certain intensities the local temperature could increase resulting in convection. In Fig.~\ref{Fig:intensity} we show how such a flow depends on the light intensity under exposure to focused UV light ($R$ = 15 $\mu$m). These results refer to the intensity varied in a broad range, from 30 mW/cm$^2$ to 10$^6$ mW/cm$^2$. The emerging flow is visualized by monitoring the passive particles's displacement (Fig.~\ref{Fig:intensity}(a-c)). Figure~\ref{Fig:intensity}(d) plots the velocity of particles, $v_p$, in the ring near the edge of the cleaned area as a function of light intensity. It is seen that the velocity of colloids, moving outwards of the beam, increases with the intensity up to 10$^4$ mW/cm$^2$. Starting from ca. 10$^5$ mW/cm$^2$ the particles move towards the light spot being driven by the temperature gradient. In the intermediate range of the intensity the counter propagating DO and convection flows are balanced and there is no particle displacement. Consequently, for all measurements of LDDO phenomena described below we use only the intensities confined in a dashed rectangle in Fig.~\ref{Fig:intensity}(d). Corresponding videos (Fig.~S1) is provided in supplemental materials.

\begin{figure}[tbp]
\begin{center}
\vspace{0.4cm} \includegraphics[width=1.0\columnwidth]{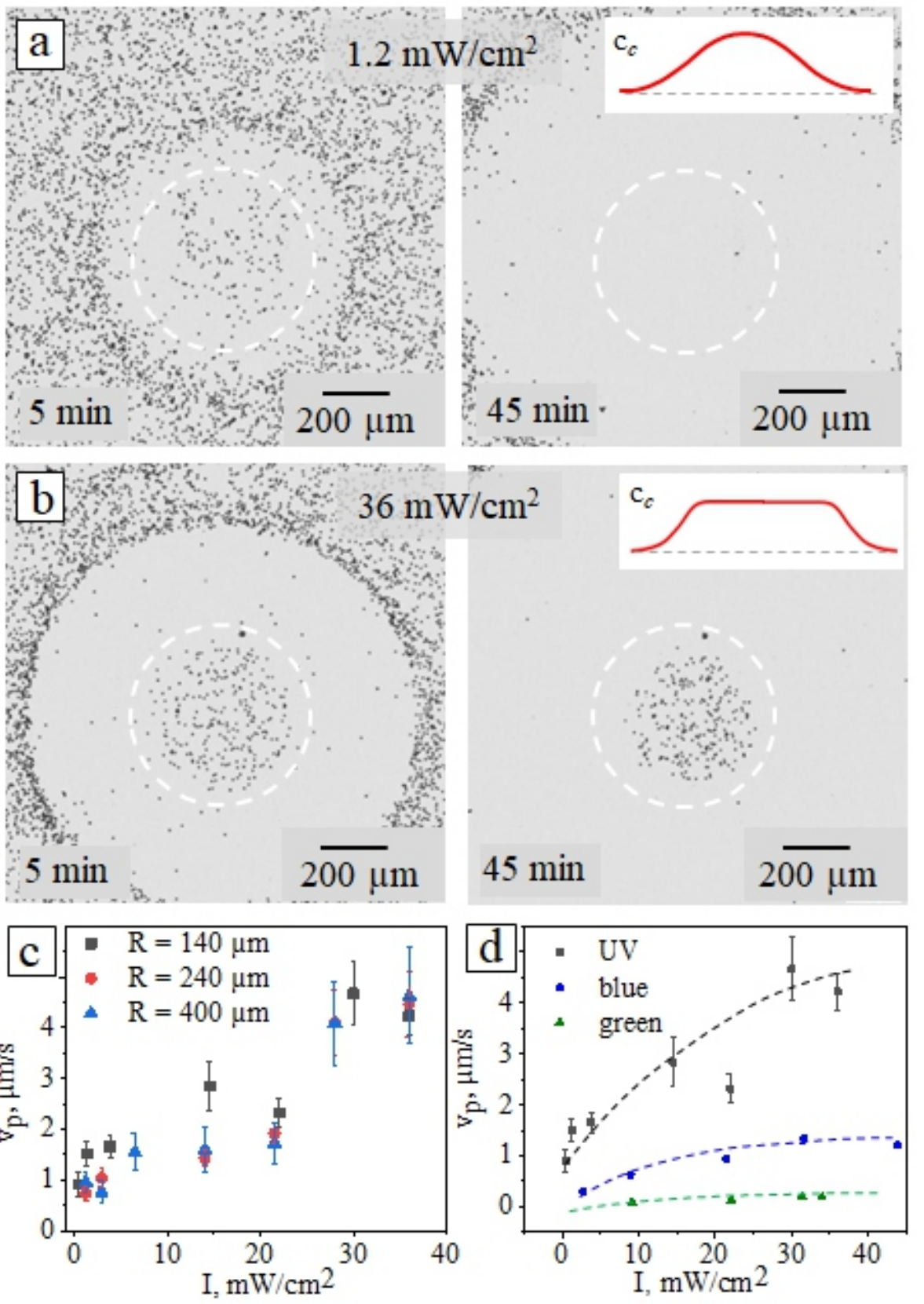}
\end{center}
\par
\vspace{-0.4cm}
\caption{ Optical micrographs of nonporous silica particles under exposure to expanded beam (that is marked by dashed white circles) of UV irradiation  ($\lambda$ = 365 nm, $R$ = 240 $\mu$m) at intensities: (a) $I$ = 1.2 mW/cm$^2$ ($t$ = 5 min and 45 min), (b) $I$ = 36 mW/cm$^2$ ($t =5$ min, 45 min). The insets illustrate the distributions of \emph{cis}-isomer concentrations. (c) Maximum velocity of the particles acquired within first minutes of irradiation as a function of light intensity for three different laser spot sizes. (d) Comparison of velocities at different intensities for different irradiation wavelengths (UV, blue and green).  The dashed lines represent a guide to the eye. The error bars are calculated as the error of the mean taken over different radius of the laser beam. }
\label{Fig:time}
\end{figure}

At large laser spot radii (greater than 100 $\mu$m) another scenario of particle motion can be observed. At small intensities, $c_c$ is everywhere nonuniform, and all particles move slowly out of the irradiated area (Fig.~\ref{Fig:time}(a)), so that we have a growing cleaned area similar to Fig.~\ref{Fig:intensity}(a). However, when $I$ becomes larger than 14 mW/cm$^2$, a large fraction of particles still move outward, but some of them remain trapped inside the central circle (Fig. \ref{Fig:time}(b)). We stress that consequent, over 4 hours, irradiation does not affect particles that are confined in a beam area, i.e. they do not leave this area  (see corresponding video in Fig.~S2 of supplemental materials). Clearly, the DO can only be generated outside of this region of a constant $c_c$, and the length scale of inhomogeneous concentration field $L$ for such a beam becomes smaller than $R$.
We measure the velocity of nonporous particles at the border of the cleaned area, which is obviously the largest velocity in the system.
Fig.~\ref{Fig:time}(c,d) shows $v_p$ as a function of irradiation intensity for different spot radii and wavelengths. Our results show that velocities increase with $I$ and are independent on $R$ (Fig.~\ref{Fig:time}(c)) which implies that $L$ also does not depend on $R$ in this case. The maximum values are attained for UV light (Fig.~\ref{Fig:time}(d)).

\begin{figure}[tbp]
\begin{center}
\vspace{0.4cm} \includegraphics[width=1.0\columnwidth]{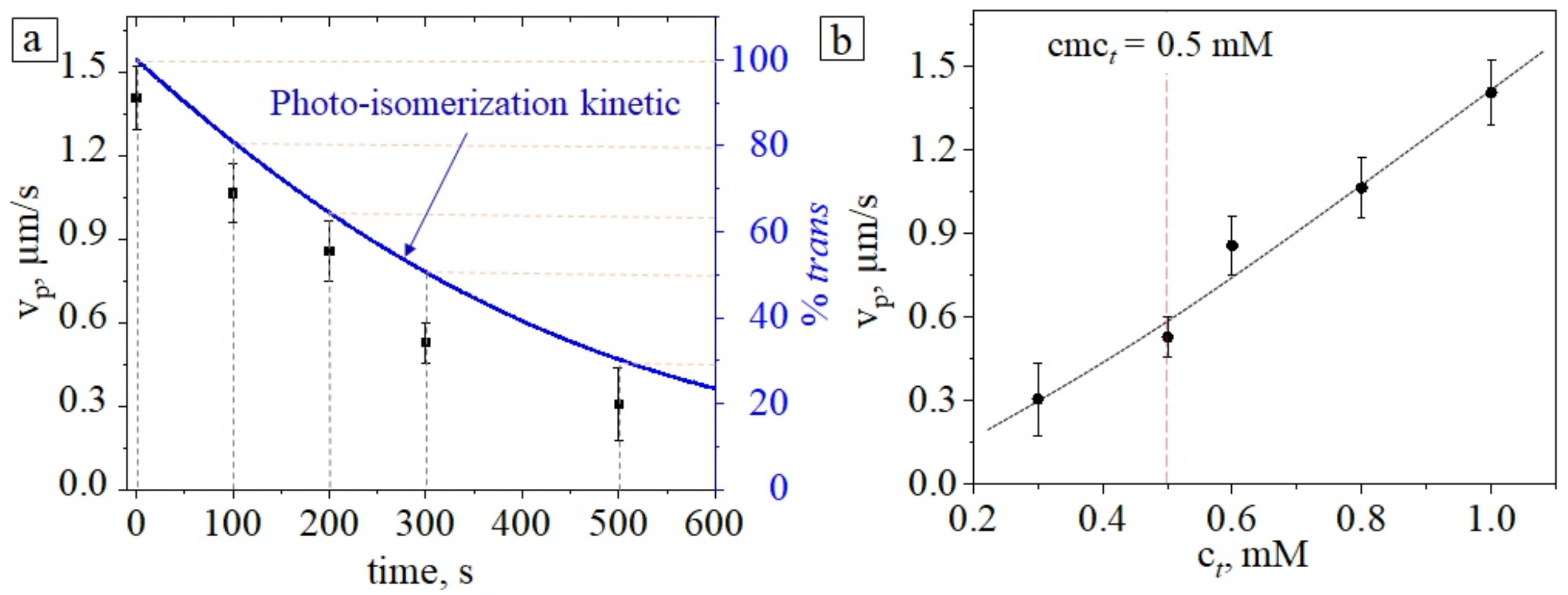}
\end{center}
\par
\vspace{-0.4cm}
\caption{(a) Maximum velocities (squares) of passive particles under irradiation with focused UV light ($\lambda$ = 375 nm, $I$ = 283 mW/cm$^2$) as a function of the time pre-irradiation by homogeneous UV light ($\lambda$ = 365 nm, I = 0.128 mW/cm$^2$). Solid curve shows the decrease in percentage of \emph{trans}-isomers in the bulk with the pre-irradiation time. (b) Particle velocity as a function of $c_t$ (circles). The black line represents a guide to the eye. Vertical dotted line indicates $cmc$.  }
\label{Fig:maxvel}
\end{figure}

The velocity of LDDO flow for particle manipulation can be adjusted by either a choice of surfactant concentration $c_0$ (see Sec.~\ref{sec:concept}) or by pre-irradiation of the sample with homogeneous UV light for a certain time. Indeed, after exposure of the surfactant solution of $c_0=1$ mM to homogeneous UV light of 0.128 mW/cm$^2$ during  100 seconds, the amount of the \emph{trans}-isomers in the bulk decreases to 80\%(see Fig.~\ref{Fig:maxvel}(a)). The particle velocity in such a pre-irradiated sample is the same as in the solution of $c_0=0.8$mM exposed to irradiation with the focused UV light (see Fig.~\ref{Fig:maxvel}(b) and S3 in supplemental materials). Correspondingly, with an increase in the pre-irradiation time, the amount of the \emph{trans}-isomers reduces (blue line in Fig.~\ref{Fig:maxvel}(a) depicts the kinetic of photo-isomerization), and the velocity drops. By an exposure of the sample to a longer wavelength, one converts the majority of surfactant molecules back to the \emph{trans}-state and thus restore the system again.
The restoring is completed when the photo-stationary state is achieved, and this takes place within few minutes of irradiation depending on the light intensity and can be at maximum of 10 minutes. Thus, one of the advantages of using LDDO particle manipulation is the ``re-charging'' of the ``battery'' within only few minutes.

\subsection{Passive nonporous versus active porous}

\subsubsection{Triggering LDDO flow by changing the wave length of light }

\begin{figure}[tbp]
\begin{center}
\vspace{0.4cm} \includegraphics[width=1.0\columnwidth]{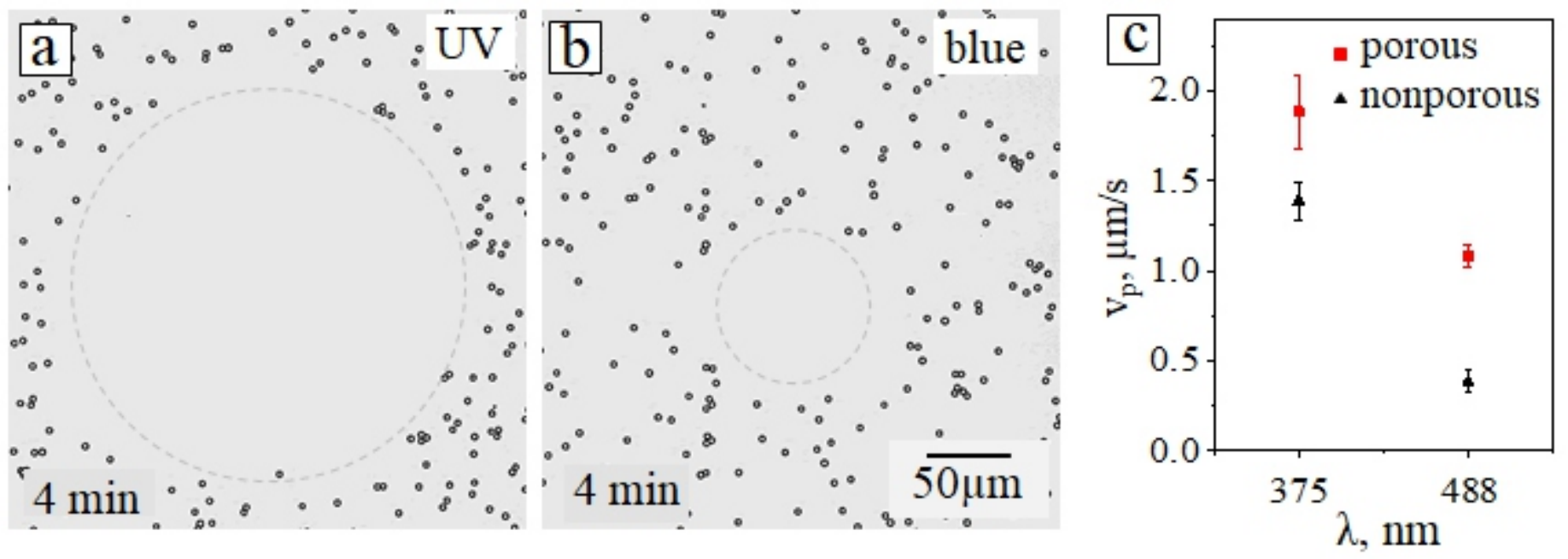}
\end{center}
\par
\vspace{-0.4cm}
\caption{(a, b) Optical micrographs of the cleaned area under irradiation with focused UV ($\lambda$ = 375 nm, $I$ =283 mW/cm$^2$) and blue light ($\lambda$ = 488 nm, $I$ =  300 mW/cm$^2$). (c) Velocities, $v_p$ of porous and nonporous particles under irradiation with 375 nm and 488  nm wavelength. }
\label{Fig:wavelength}
\end{figure}

We next examine the displacement of a mixture of passive nonporous and active porous particles illuminated by a focused light. The results are shown in  Fig.~\ref{Fig:wavelength}.
The direction and the velocity of the LDDO flow depend sensitively on the local distribution of the \emph{trans}- and \emph{cis}-isomers. Namely, the flow is directed out of the area of the increased \emph{cis}-concentration  into the \emph{trans}-enriched area. The corresponding concentration gradients can be generated in different ways. For instance, under UV exposure more than 90\% of the surfactant is isomerized to the \emph{cis}-state.
Also, a smaller amount of surfactant molecules in the \emph{cis}-state is formed under irradiation with blue and green lights (see Fig.~\ref{Fig:wavelength} (c)).  In all these cases the outwards flow is generated, and its velocities and radii of the cleaned (i.e. free of particles) area depend on the wavelength (see Fig.~\ref{Fig:wavelength} (a,b,d)). An important observation is that the velocities of the porous and nonporous particles shown in Fig.~\ref{Fig:wavelength}(d) are different. Namely, porous particles are generally faster than similar impermeable colloids. This is exactly what we have suggested in Sec.~\ref{sec:theory_active} by arguing that active particles should have a smaller contact friction on the bottom wall.
This experimental result  requires further theoretical studies, which are beyond the scope of the present paper.


\begin{figure}[tbp]
\begin{center}
\vspace{0.4cm} \includegraphics[width=1.0\columnwidth]{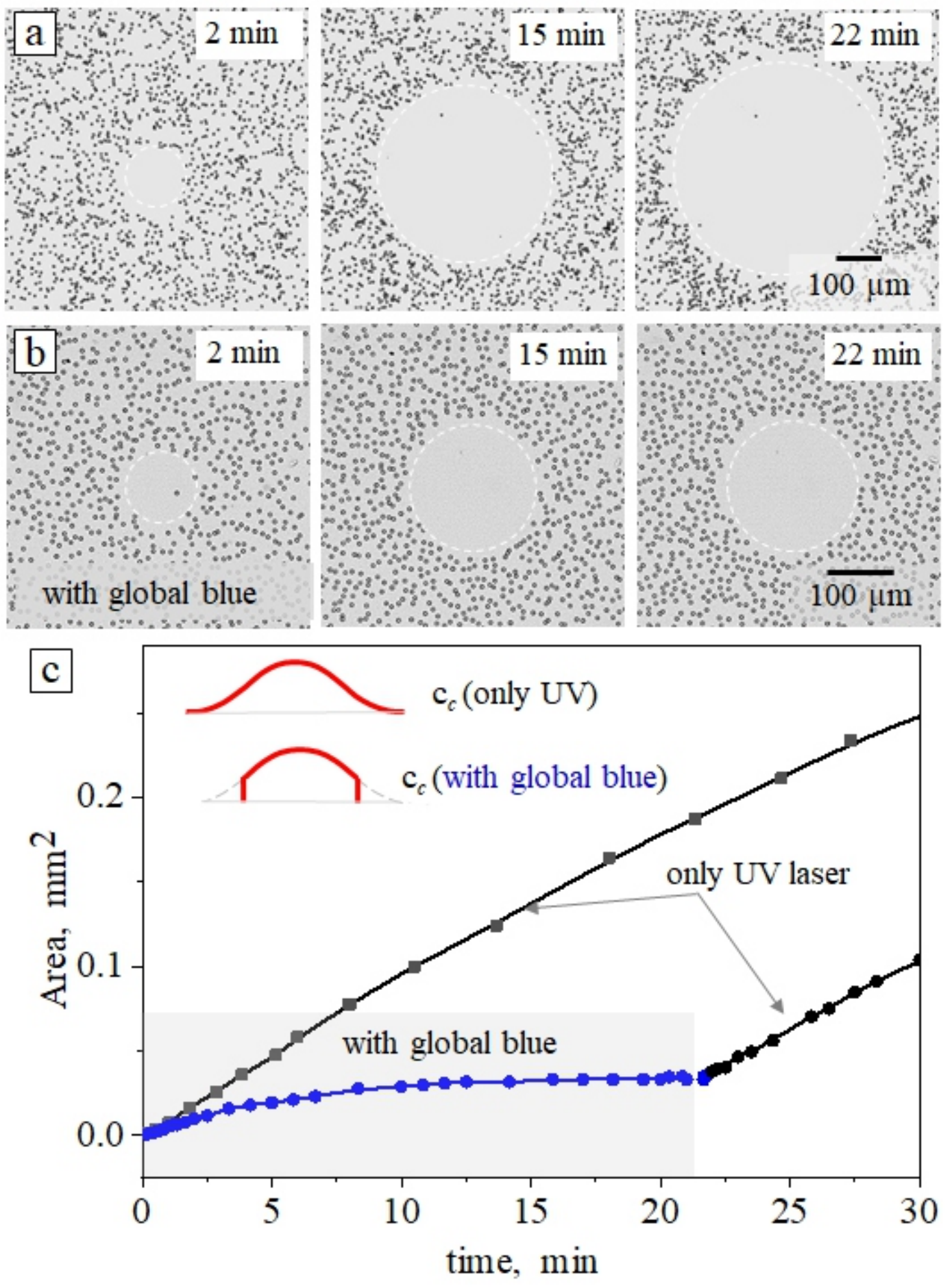}
\end{center}
\par
\vspace{-0.4cm}
\caption{Optical micrographs of passive (nonporous) silica particles at several irradiation time (of 2, 15, and 22 min) exposed to (a) focused UV irradiation ($\lambda$ = 375 nm, $I$ = 289 mW/cm$^2$) and (b) simultaneous irradiation with focused UV and homogeneous blue light ($\lambda$ = 455 nm, $I$ = 3.4 mW/cm$^2$). (c) Corresponding clean area as a function of time. The inset illustrates the distribution of \emph{cis}-isomer concentration without (upper curve) and with (lower curve) homogeneous blue irradiation.  }
\label{Fig:ir_time}
\end{figure}

We next address another important question of how long does the LDDO flow sustain under irradiation with a focused light? Fig.~\ref{Fig:ir_time}(a) illustrates an extension of  the cleaned of nonporous particles area under the irradiation with the focused UV light at different moments of time. The area of the cleaned region grows with time strictly monotonically (see uppermost curve in Fig.~\ref{Fig:ir_time}(c) and also Fig. S4 in supplemental materials for the growth during 5 hours).
However, when the same focused UV light is applied together with  an homogeneous blue light, the growth of the cleaned area  stops already after 10 minutes as seen in Fig.~\ref{Fig:ir_time}(b), but by switching off the blue light one can restore the growth of the cleaned area. As a result, the area of the cleaned region augments weakly monotonically (see lowermost curve in Fig.~\ref{Fig:ir_time}(c) and shows a distinct plateau at the time range from ca. 10 to 20 min).
These results suggest that one can easily control the size of the cleaned area simply by switching on and off the homogeneous blue light while continuously irradiating with focused UV beam (see supplemental materials for a movie in Fig.~S5).

This behavior is likely related to the restricted concentration gradient of the \emph{cis}-isomers in the irradiated area. Indeed, under a continuous exposure to the focused UV beam with a switch-off blue light, the concentration gradient is defined by the diffusion of the \emph{cis}-isomers out of the irradiated area (see inset in Fig.~\ref{Fig:surfactant}(c)). In the case of the superimposed blue light, the gradient is cut by the continuous photo-isomerization out of the UV irradiated area, so that the \emph{cis}-isomers produced in the focused light escape outside  and isomerizes back to the \emph{trans}- one. The size of the cleaned area is then defined by the ratio of the blue/UV light intensities since the amounts of the \emph{trans}- and \emph{cis}- isomers vary significantly.


\begin{figure}[tbp]
\begin{center}
\vspace{0.4cm} \includegraphics[width=1.0\columnwidth]{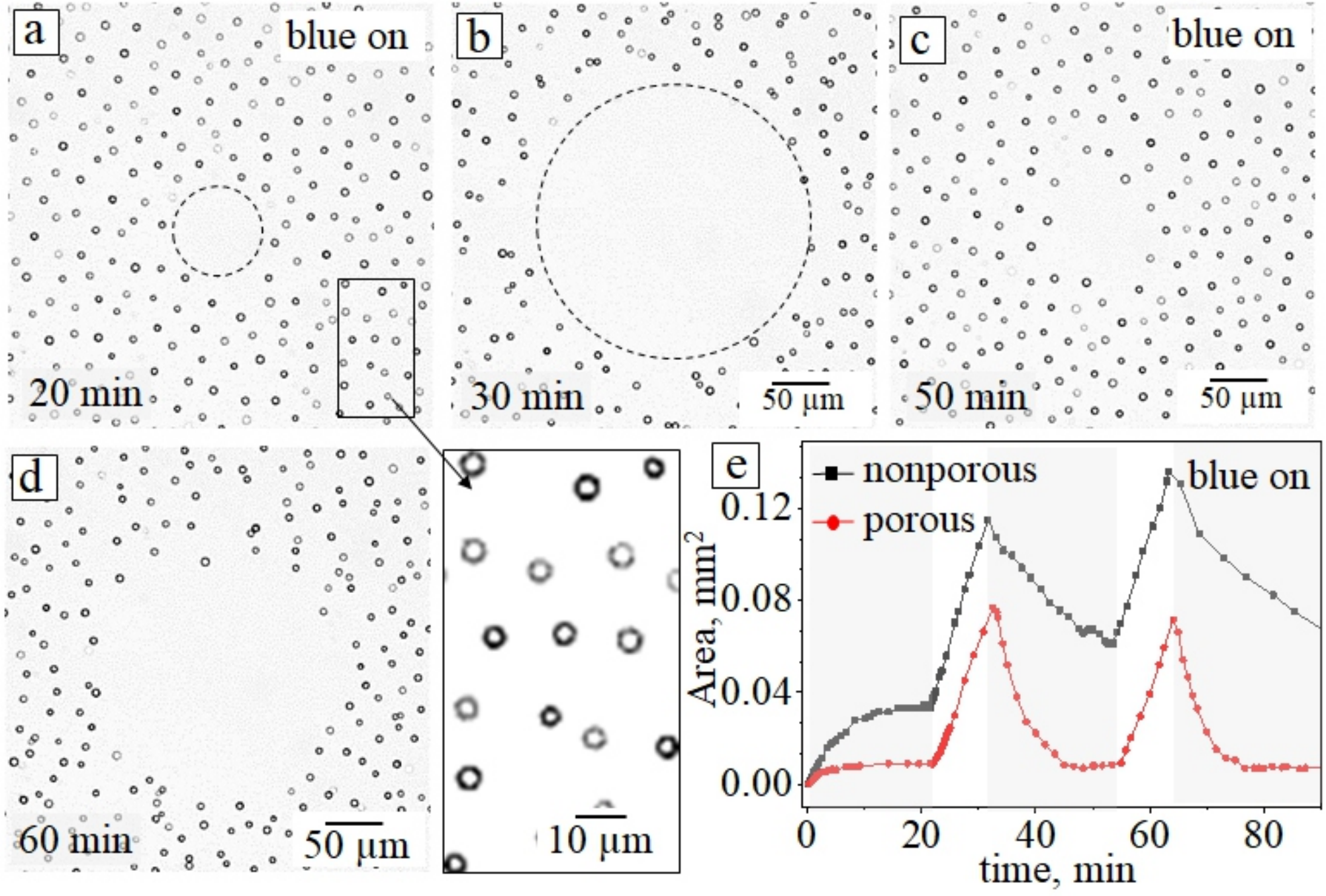}
\end{center}
\par
\vspace{-0.4cm}
\caption{(a-d) Optical micrographs of the porous particles exposed to irradiation with focused UV light ($\lambda$ = 375 nm, $I$ = 289 mW/cm$^2$) and additional blue irradiation ($\lambda$ = 455 nm, I= 3.4 mW/cm$^2$) periodically switched on (a, c). Enlargement of the selected area in (a) shows separated  due to DO repulsion active colloids. (e) Dependence of the cleaned area  for nonporous (black curve) and porous (red curve) particles on the irradiation time when the blue light is periodically applied (marked in grey). }
\label{Fig:porous_focused}
\end{figure}

Similar response is observed in the case of active porous particles (see Fig.~\ref{Fig:porous_focused}).
We remark that the size of the cleaned area is always smaller than for nonporous particles. At first sight this is somewhat surprising, since, as discussed above, the porous particles should generally move faster under irradiation with light of one wavelength, either UV or blue (see Fig.~\ref{Fig:wavelength}(c)). We recall, however, that in this particular experiment we have a combination of the two wavelengths, homogeneous blue and focused UV. In this case the cleaned area for active particles decreases faster when the blue light is switched on (see Fig.~\ref{Fig:porous_focused}(e) and corresponding video in Figure S6 of supplemental materials). This effect is due to particle mutual DO repulsion (described in Sec.~\ref{sec:theory_active} and illustrated in Figs.~\ref{Fig:porous_focused}(a,c)) outside the cleaned area.

\subsection{Controlled particle manipulation by optically triggered restrictions}

\begin{figure}[tbp]
\begin{center}
\vspace{0.4cm} \includegraphics[width=1.0\columnwidth]{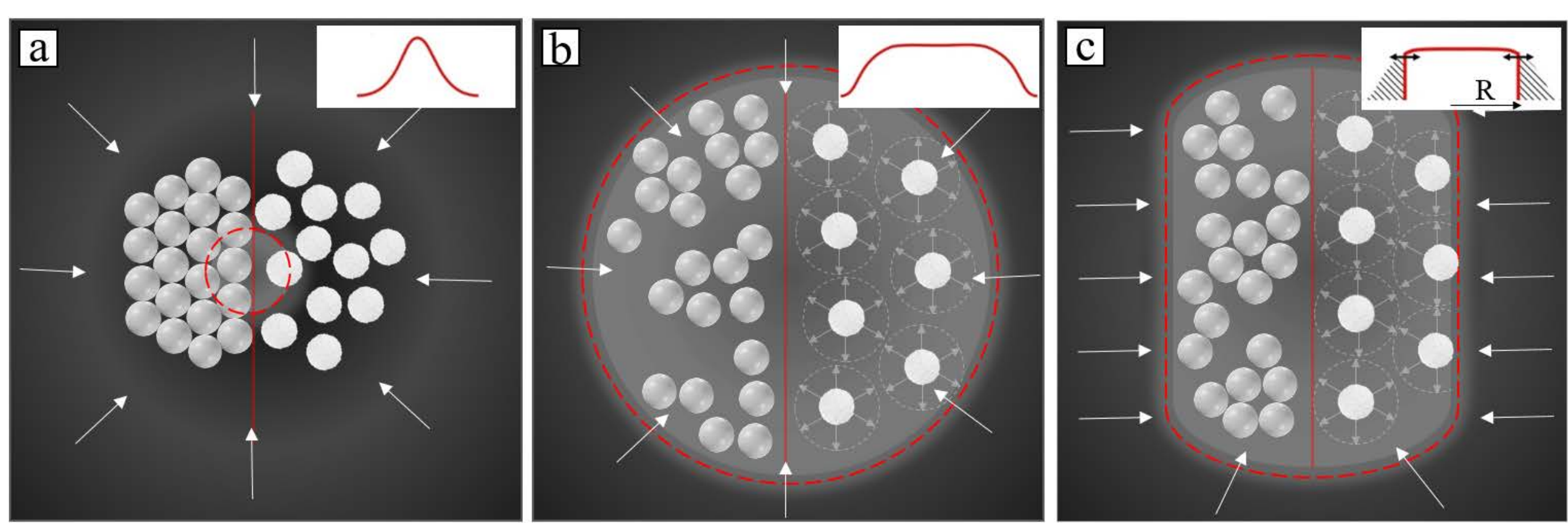}
\end{center}
\par
\vspace{-0.4cm}
\caption{(a, b) Particles confined in the shaped into a round form Gaussian and flat-top beams. Grey spheres on the left correspond to nonporous passive particles. White  spheres (decorated with small radial arrows  indicating a DO repulsion) on the right plot active porous particles. (c) The same as in (b), but the flat-top beam is of  the elongated, nearly rectangular, shape and its ideality can be controlled by changing the radius of the beam at certain frequency.  White arrows show the direction of LDDO flow,
red dashed curves  mark boundaries of the light spot.    }
\label{Fig:shape}
\end{figure}

We are now on a position to  propose the recipes of selecting and further confining a subset of particles in a confined region, without using a chamber with real walls. 

Fig.~\ref{Fig:shape} illustrates how the expanding of the laser spot or changing its parameters, such as the shape and/or intensity profile, could potentially  manipulate arrested  nonporous (grey spheres) or active porous particles (white spheres). For example, a Gaussian beam should tightly gather particles \cite{Arya.P:2020c} as seen in Fig.~\ref{Fig:shape}(a), but using flat-top beams of circular or rectangular shapes would almost likely lead to more dilute ensembles, with much larger separations between active particles (Fig.~\ref{Fig:shape}(b,c)). Moreover, the confinement (and the ideality of the flat-top beam) could be dynamically controlled by the changing the radius of the beam at certain frequency (Fig.~\ref{Fig:shape}(c)). Thus, since the size of a confined area and the uniformity of the illumination  can be easily tuned,
one may generate situations from rather dilute ``gases'' of colloids to almost crystalline arrangements. Below we present  some specimen experimental results illustrating the power and variability of this approach.

\begin{figure}[tbp]
\begin{center}
\vspace{0.4cm} \includegraphics[width=1.0\columnwidth]{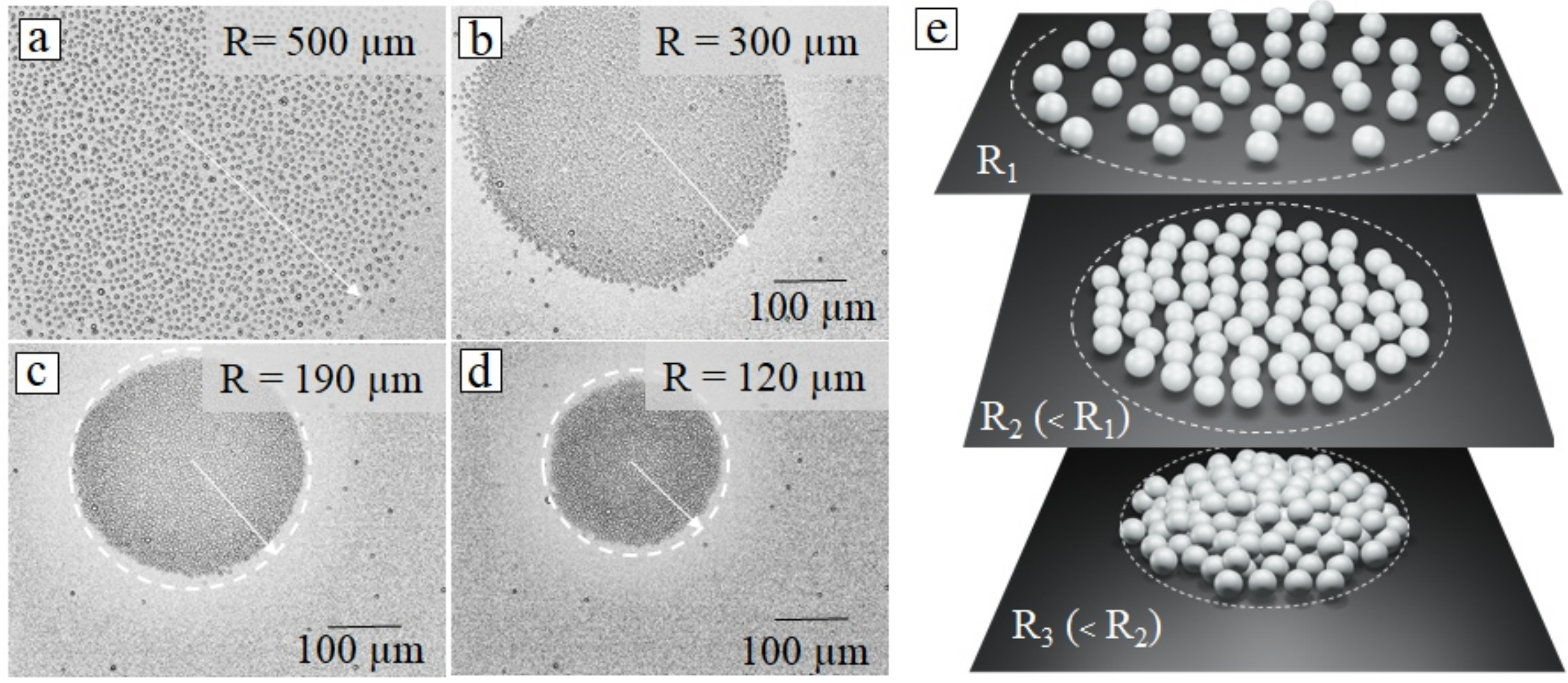}
\end{center}
\par
\vspace{-0.4cm}
\caption{(a-d) Optical micrographs showing the collection of nonporous particles by changing the size of the circular laser spot ($\lambda$=491 nm, $I$ = 234 mW/cm$^2$). (e) Schematic of the process. }
\label{Fig:shape2}
\end{figure}

\begin{figure}[tbp]
\begin{center}
\vspace{0.4cm} \includegraphics[width=1.0\columnwidth]{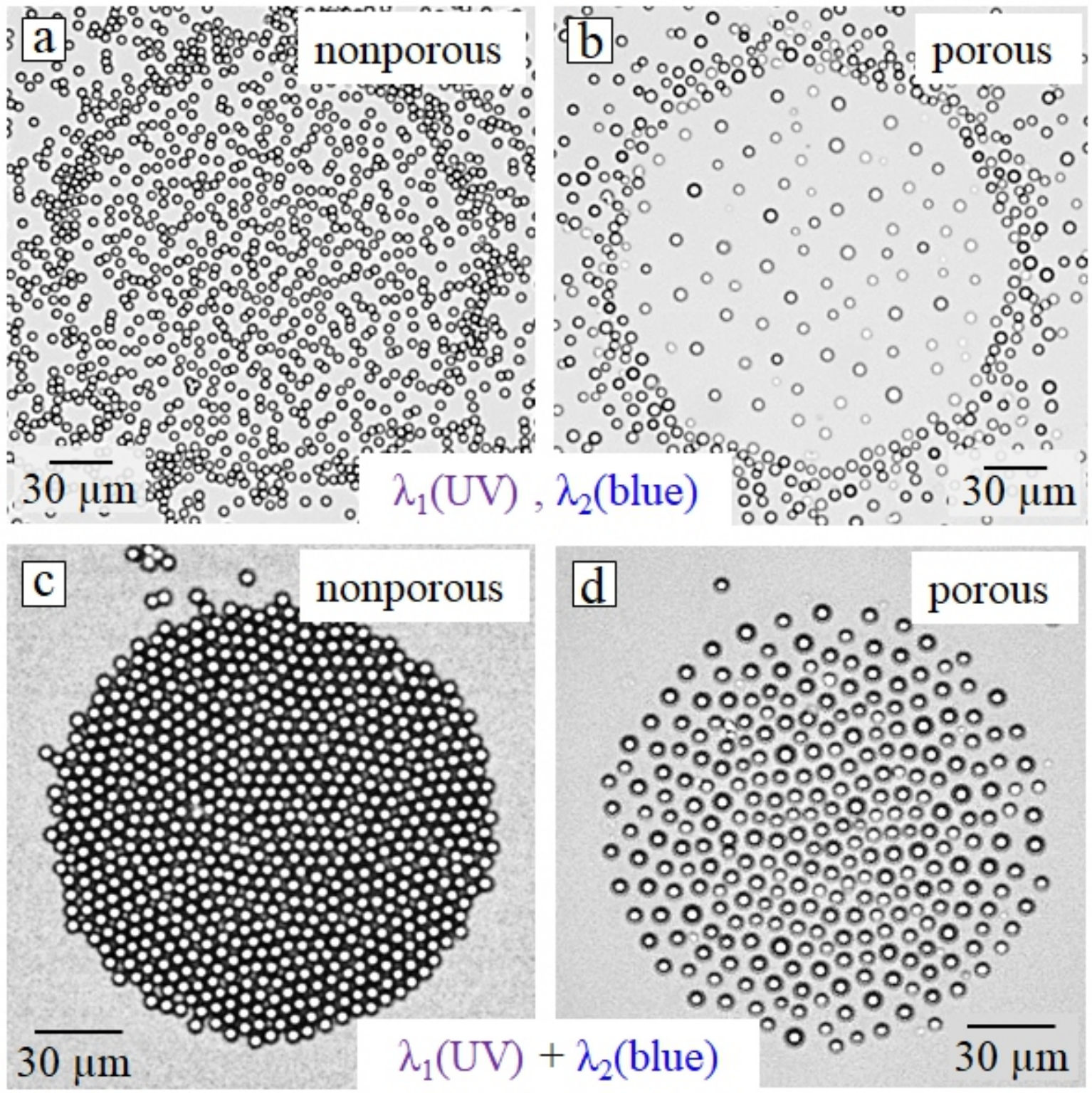}
\end{center}
\par
\vspace{-0.4cm}
\caption{Optical micrographs of nonporous (left) and porous active (right) silica particles under different irradiation conditions. (a, b) Consecutive irradiation with homogeneous UV ($\lambda$ = 365 nm, $I$= 5.5 mW/cm$^2$ ) followed by exposure to focused blue light ($R$ = 280 $\mu$m, $\lambda$ = 455 nm, $I$ = 17.9 mW/cm$^2$). (c, d) Simultaneous irradiation with homogeneous UV ($\lambda$ = 365 nm, $I$= 1.4 mW/cm$^2$ ) and focused blue light ($R$ = 160 $\mu$m, $\lambda$ = 455 nm, $I$ = 234 mW/cm$^2$).    }
\label{Fig:irradiation}
\end{figure}

Fig.~\ref{Fig:shape2} shows that upon reducing the radius of the circular laser spot, $R$, from 500 $\mu$m down to 190 $\mu$m, one can generate a rapid change of the particle surface density, from a dilute state to a densely packed monolayer (Fig.~\ref{Fig:shape2}(a, b, c)). On reducing $R$ further one can even force the particles to form a multi-layer  crystal structure (see Fig.~\ref{Fig:shape2}(d)). Note that in this specimen example we increase the density of the confined particles in 17 times. We remark and stress that the process is completely reversible.

In the case of  irradiation with homogeneous UV followed by focused blue light, i.e. when the blue light is applied to a \emph{cis}-enriched solution, the outer particles are concentrated at the boarder of the irradiated spot forming a densely packed shell. The ``incapsulated'' inner particles, i.e. the ones in the confined area, randomly move if they are nonporous (Fig.~\ref{Fig:irradiation}(a)), but when porous, a kind of lattice of well-separated (due to the DO repulsion) particles is formed
(Fig.~\ref{Fig:irradiation}(b)). When these  systems  are irradiated by the combined light of two wavelengths, namely, when the UV light is homogeneously exposed over the sample, while the selected area is illuminated by blue light, steady configurations of particles become different from described above. In this case, all particles are collected to the focused spot area, and their package becomes denser then in Figs.~\ref{Fig:irradiation}(a, b). The passive particles form a crystalline structure (Fig.~\ref{Fig:irradiation}(c)). The active ones remain separated due to the DO repulsion (Fig.~\ref{Fig:irradiation}(d)), but the interparticle distance becomes quite small.

\begin{figure}[tbp]
\begin{center}
\vspace{0.4cm} \includegraphics[width=1.0\columnwidth]{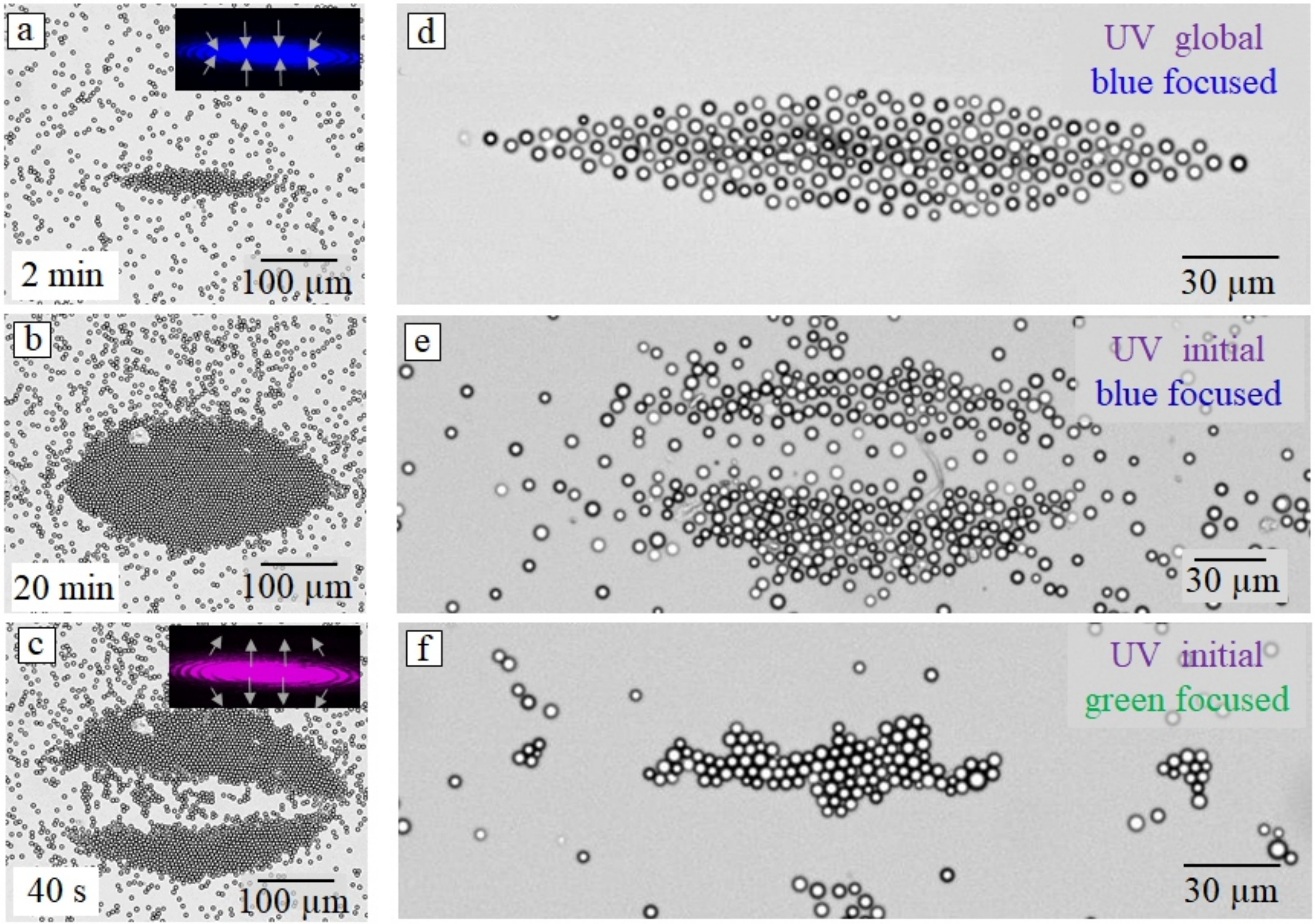}
\end{center}
\par
\vspace{-0.4cm}
\caption{(a,b,c) Optical micrographs of nonporous silica particles during exposure to laser spot of an elongated shape of either blue or UV light. Insets in (a) and (c) show the beam shape for blue and UV light, respectively, with arrows indicating the direction of the DO flow. (d,e,f) Optical micrographs of porous silica colloids during irradiation with modulated laser spot: (d) blue light with additional homogeneous irradiation with UV light,  (e) blue light applied to pre-irradiated sample with homogeneous UV light, (f) green light applied to pre-irradiated sample with UV light.   }
\label{Fig:elongated}
\end{figure}

Fig.~\ref{Fig:elongated} demonstrates that one can easily ``cut'' an optically confined dispersion of sedimented particles, by acting as optical ``scissors''. When particles are passive nonporous (Fig.~\ref{Fig:elongated}(a)), the irradiation of the \emph{cis}-enriched solution with modulated blue beam of elliptic shape ($R_1= 215$ $\mu$m, $R_2= 36$ $\mu$m) leads to a formation of the elongated, densely packed cluster (Fig.~\ref{Fig:elongated}(b)). By applying then the UV light this cluster of nonporous particles can be splitted by two parts (see corresponding video in Fig. S7 of supplemental materials). Porous particles can be manipulated similarly, and we illustrate this by using other illumination conditions. When the particles are irradiated by focused blue light together with homogeneous UV light, the elongated clusters arise as seen in Fig.~\ref{Fig:elongated}(d). If we apply homogeneous UV pre-irradiation and then superimpose focused blue light, one can easily cut the cluster of particles (see Fig.~\ref{Fig:elongated}(e)). However, an homogeneous UV pre-irradiation with the consequent illumination with focused green light lead to a tightly arranged elongated ensemble of the porous particles resembling the situation with nonporous colloids   (Fig.~\ref{Fig:elongated}(f)). Note that the generated patterns can again be easily reversed by tuning the  parameters of the laser beam (see corresponding videos in Figure S8 of supplemental materials).


\section{Conclusion}\label{Sec:conclusion}

In our work we have demonstrated that a combination of two distinct primitive scenarios  of the light-induced diffusio-osmosis (LDDO) can be exploited to generate many  non-trivial dynamic configurations of particle ensembles at a solid wall, at once, seamlessly and quickly.
For instance, it can be used to  create the  restricted, but ``fenceless'', areas of tunable size, where the motion of passive and active particles can be studied and/or remotely controlled. Importantly,  we can easily confine a subset of particles simply by using the laser spot, i.e. without introducing them into a chamber with physical walls. We also note that by simply varying the radiation intensity, focusing or wavelength is sufficient to alter the diffusio-osmotic flow  and thus to induce various patterns of colloid particles at the bottom wall, from rather dilute ``gases'' of colloids to almost crystalline, densely packed structures.

One of the main experimental results of our work is the difference in behaviour of nonporous and porous particles.  The trends that have been observed are
consistent with our simple theoretical models that treat nonporous particles as passive, and porous - as actively participating in the mechanism of the flow generation.

Our strategy can be employed  to manipulate more complex particles, such as asymmetric Janus ones that contain active and passive parts. A systematic study of their manipulation in the spirit of the present work appears to be very timely and would
constitute its significant extension.

\section*{Acknowledgments}

This research is supported by the Priority Program 1726 ``Microswimmers-From
Single Particle Motion to Collective Behaviour'' of the German Research
Foundation (SA1657/9-2 and VI 243/4-2), the International Max Planck
Research School on Multiscale Bio-Systems (IMPRS), and the Russian Ministry
of Education and Science. We thank A. Kopyshev for the support in
preparation of figures.

\section*{Author contribution statement}

S.S. and O.I.V. designed and supervised the project and wrote the
manuscript.  P.A., M.U., J.J. and D.F. performed the device fabrication,
measurements, and analysis of the experimental data. N.L. synthesized the
photosensitive surfactant. E.S.A. and O.I.V. developed the theory. All the
authors were involved in the preparation of the manuscript. All the authors
have read and approved the final manuscript.


\begin{thebibliography}{46}%
\makeatletter
\providecommand \@ifxundefined [1]{%
 \@ifx{#1\undefined}
}%
\providecommand \@ifnum [1]{%
 \ifnum #1\expandafter \@firstoftwo
 \else \expandafter \@secondoftwo
 \fi
}%
\providecommand \@ifx [1]{%
 \ifx #1\expandafter \@firstoftwo
 \else \expandafter \@secondoftwo
 \fi
}%
\providecommand \natexlab [1]{#1}%
\providecommand \enquote  [1]{``#1''}%
\providecommand \bibnamefont  [1]{#1}%
\providecommand \bibfnamefont [1]{#1}%
\providecommand \citenamefont [1]{#1}%
\providecommand \href@noop [0]{\@secondoftwo}%
\providecommand \href [0]{\begingroup \@sanitize@url \@href}%
\providecommand \@href[1]{\@@startlink{#1}\@@href}%
\providecommand \@@href[1]{\endgroup#1\@@endlink}%
\providecommand \@sanitize@url [0]{\catcode `\\12\catcode `\$12\catcode
  `\&12\catcode `\#12\catcode `\^12\catcode `\_12\catcode `\%12\relax}%
\providecommand \@@startlink[1]{}%
\providecommand \@@endlink[0]{}%
\providecommand \url  [0]{\begingroup\@sanitize@url \@url }%
\providecommand \@url [1]{\endgroup\@href {#1}{\urlprefix }}%
\providecommand \urlprefix  [0]{URL }%
\providecommand \Eprint [0]{\href }%
\providecommand \doibase [0]{http://dx.doi.org/}%
\providecommand \selectlanguage [0]{\@gobble}%
\providecommand \bibinfo  [0]{\@secondoftwo}%
\providecommand \bibfield  [0]{\@secondoftwo}%
\providecommand \translation [1]{[#1]}%
\providecommand \BibitemOpen [0]{}%
\providecommand \bibitemStop [0]{}%
\providecommand \bibitemNoStop [0]{.\EOS\space}%
\providecommand \EOS [0]{\spacefactor3000\relax}%
\providecommand \BibitemShut  [1]{\csname bibitem#1\endcsname}%
\let\auto@bib@innerbib\@empty
\bibitem [{\citenamefont {Zhang}\ \emph {et~al.}(2020)\citenamefont {Zhang},
  \citenamefont {Bu}, \citenamefont {Yip}, \citenamefont {Liang},\ and\
  \citenamefont {Ho}}]{Zhang.H:2020}%
  \BibitemOpen
  \bibfield  {author} {\bibinfo {author} {\bibfnamefont {H.}~\bibnamefont
  {Zhang}}, \bibinfo {author} {\bibfnamefont {X.}~\bibnamefont {Bu}}, \bibinfo
  {author} {\bibfnamefont {S.}~\bibnamefont {Yip}}, \bibinfo {author}
  {\bibfnamefont {X.}~\bibnamefont {Liang}}, \ and\ \bibinfo {author}
  {\bibfnamefont {J.~C.}\ \bibnamefont {Ho}},\ }\href@noop {} {\bibfield
  {journal} {\bibinfo  {journal} {Adv. Intell. Syst.}\ }\textbf {\bibinfo
  {volume} {2}},\ \bibinfo {pages} {1900085} (\bibinfo {year}
  {2020})}\BibitemShut {NoStop}%
\bibitem [{\citenamefont {Wang}\ and\ \citenamefont {Qi}(2019)}]{Wang.W:2019}%
  \BibitemOpen
  \bibfield  {author} {\bibinfo {author} {\bibfnamefont {W.}~\bibnamefont
  {Wang}}\ and\ \bibinfo {author} {\bibfnamefont {L.}~\bibnamefont {Qi}},\
  }\href@noop {} {\bibfield  {journal} {\bibinfo  {journal} {Adv. Funct.
  Mater.}\ }\textbf {\bibinfo {volume} {29}},\ \bibinfo {pages} {1807275}
  (\bibinfo {year} {2019})}\BibitemShut {NoStop}%
\bibitem [{\citenamefont {Bian}\ \emph {et~al.}(2020)\citenamefont {Bian},
  \citenamefont {Sun}, \citenamefont {Cai}, \citenamefont {Wang}, \citenamefont
  {Wang},\ and\ \citenamefont {Zhao}}]{Bian.F:2020}%
  \BibitemOpen
  \bibfield  {author} {\bibinfo {author} {\bibfnamefont {F.}~\bibnamefont
  {Bian}}, \bibinfo {author} {\bibfnamefont {L.}~\bibnamefont {Sun}}, \bibinfo
  {author} {\bibfnamefont {L.}~\bibnamefont {Cai}}, \bibinfo {author}
  {\bibfnamefont {Y.}~\bibnamefont {Wang}}, \bibinfo {author} {\bibfnamefont
  {Y.}~\bibnamefont {Wang}}, \ and\ \bibinfo {author} {\bibfnamefont
  {Y.}~\bibnamefont {Zhao}},\ }\href@noop {} {\bibfield  {journal} {\bibinfo
  {journal} {Small}\ }\textbf {\bibinfo {volume} {16}},\ \bibinfo {pages}
  {1903931} (\bibinfo {year} {2020})}\BibitemShut {NoStop}%
\bibitem [{\citenamefont {Wong}\ \emph {et~al.}(2011)\citenamefont {Wong},
  \citenamefont {Kang}, \citenamefont {Tang}, \citenamefont {Smythe},
  \citenamefont {Hatton}, \citenamefont {Grinthal},\ and\ \citenamefont
  {Aizenberg}}]{Wong.T:2011}%
  \BibitemOpen
  \bibfield  {author} {\bibinfo {author} {\bibfnamefont {T.}~\bibnamefont
  {Wong}}, \bibinfo {author} {\bibfnamefont {S.~H.}\ \bibnamefont {Kang}},
  \bibinfo {author} {\bibfnamefont {S.~K.~Y.}\ \bibnamefont {Tang}}, \bibinfo
  {author} {\bibfnamefont {E.~J.}\ \bibnamefont {Smythe}}, \bibinfo {author}
  {\bibfnamefont {B.~D.}\ \bibnamefont {Hatton}}, \bibinfo {author}
  {\bibfnamefont {A.}~\bibnamefont {Grinthal}}, \ and\ \bibinfo {author}
  {\bibfnamefont {J.}~\bibnamefont {Aizenberg}},\ }\href@noop {} {\bibfield
  {journal} {\bibinfo  {journal} {Nature}\ }\textbf {\bibinfo {volume} {477}},\
  \bibinfo {pages} {443} (\bibinfo {year} {2011})}\BibitemShut {NoStop}%
\bibitem [{\citenamefont {Rogers}\ \emph {et~al.}(2010)\citenamefont {Rogers},
  \citenamefont {Someya},\ and\ \citenamefont {Huang}}]{Rogers.J:2010}%
  \BibitemOpen
  \bibfield  {author} {\bibinfo {author} {\bibfnamefont {J.~A.}\ \bibnamefont
  {Rogers}}, \bibinfo {author} {\bibfnamefont {T.}~\bibnamefont {Someya}}, \
  and\ \bibinfo {author} {\bibfnamefont {Y.}~\bibnamefont {Huang}},\
  }\href@noop {} {\bibfield  {journal} {\bibinfo  {journal} {Science}\ }\textbf
  {\bibinfo {volume} {327}},\ \bibinfo {pages} {1603} (\bibinfo {year}
  {2010})}\BibitemShut {NoStop}%
\bibitem [{\citenamefont {Fennimore}\ \emph {et~al.}(2003)\citenamefont
  {Fennimore}, \citenamefont {Yuzvinsky}, \citenamefont {Han}, \citenamefont
  {Fuhrer}, \citenamefont {Cumings},\ and\ \citenamefont
  {Zettl}}]{Fennimore.A:2003}%
  \BibitemOpen
  \bibfield  {author} {\bibinfo {author} {\bibfnamefont {A.~M.}\ \bibnamefont
  {Fennimore}}, \bibinfo {author} {\bibfnamefont {T.~D.}\ \bibnamefont
  {Yuzvinsky}}, \bibinfo {author} {\bibfnamefont {W.-Q.}\ \bibnamefont {Han}},
  \bibinfo {author} {\bibfnamefont {M.~S.}\ \bibnamefont {Fuhrer}}, \bibinfo
  {author} {\bibfnamefont {J.}~\bibnamefont {Cumings}}, \ and\ \bibinfo
  {author} {\bibfnamefont {A.}~\bibnamefont {Zettl}},\ }\href@noop {}
  {\bibfield  {journal} {\bibinfo  {journal} {Nature}\ }\textbf {\bibinfo
  {volume} {424}},\ \bibinfo {pages} {408} (\bibinfo {year}
  {2003})}\BibitemShut {NoStop}%
\bibitem [{\citenamefont {Liljestr{\"o}m}\ \emph {et~al.}(2019)\citenamefont
  {Liljestr{\"o}m}, \citenamefont {Chen}, \citenamefont {Dommersnes},
  \citenamefont {Fossum},\ and\ \citenamefont
  {Gr{\"o}schel}}]{Liljestrom.V:2019}%
  \BibitemOpen
  \bibfield  {author} {\bibinfo {author} {\bibfnamefont {V.}~\bibnamefont
  {Liljestr{\"o}m}}, \bibinfo {author} {\bibfnamefont {C.}~\bibnamefont
  {Chen}}, \bibinfo {author} {\bibfnamefont {P.}~\bibnamefont {Dommersnes}},
  \bibinfo {author} {\bibfnamefont {J.~O.}\ \bibnamefont {Fossum}}, \ and\
  \bibinfo {author} {\bibfnamefont {A.~H.}\ \bibnamefont {Gr{\"o}schel}},\
  }\href@noop {} {\bibfield  {journal} {\bibinfo  {journal} {Curr. Opinion
  Colloid Interface Sci.}\ }\textbf {\bibinfo {volume} {40}},\ \bibinfo {pages}
  {25} (\bibinfo {year} {2019})}\BibitemShut {NoStop}%
\bibitem [{\citenamefont {Fu}\ \emph {et~al.}(2016)\citenamefont {Fu},
  \citenamefont {Xiao}, \citenamefont {Feoktystov}, \citenamefont {Pipich},
  \citenamefont {Appavou}, \citenamefont {Su}, \citenamefont {Feng},
  \citenamefont {Jin},\ and\ \citenamefont {Brückel}}]{Fu.Z:2016}%
  \BibitemOpen
  \bibfield  {author} {\bibinfo {author} {\bibfnamefont {Z.}~\bibnamefont
  {Fu}}, \bibinfo {author} {\bibfnamefont {Y.}~\bibnamefont {Xiao}}, \bibinfo
  {author} {\bibfnamefont {A.}~\bibnamefont {Feoktystov}}, \bibinfo {author}
  {\bibfnamefont {V.}~\bibnamefont {Pipich}}, \bibinfo {author} {\bibfnamefont
  {M.~S.}\ \bibnamefont {Appavou}}, \bibinfo {author} {\bibfnamefont
  {Y.}~\bibnamefont {Su}}, \bibinfo {author} {\bibfnamefont {E.}~\bibnamefont
  {Feng}}, \bibinfo {author} {\bibfnamefont {W.}~\bibnamefont {Jin}}, \ and\
  \bibinfo {author} {\bibfnamefont {T.}~\bibnamefont {Brückel}},\ }\href@noop
  {} {\bibfield  {journal} {\bibinfo  {journal} {Nanoscale}\ }\textbf {\bibinfo
  {volume} {8}},\ \bibinfo {pages} {18541} (\bibinfo {year}
  {2016})}\BibitemShut {NoStop}%
\bibitem [{\citenamefont {Crassous}\ \emph {et~al.}(2014)\citenamefont
  {Crassous}, \citenamefont {Mihut}, \citenamefont {Wernersson}, \citenamefont
  {Pfleiderer}, \citenamefont {Vermant}, \citenamefont {Linse},\ and\
  \citenamefont {Schurtenberger}}]{Crassous.J:2014}%
  \BibitemOpen
  \bibfield  {author} {\bibinfo {author} {\bibfnamefont {J.~J.}\ \bibnamefont
  {Crassous}}, \bibinfo {author} {\bibfnamefont {A.~M.}\ \bibnamefont {Mihut}},
  \bibinfo {author} {\bibfnamefont {E.}~\bibnamefont {Wernersson}}, \bibinfo
  {author} {\bibfnamefont {P.}~\bibnamefont {Pfleiderer}}, \bibinfo {author}
  {\bibfnamefont {J.}~\bibnamefont {Vermant}}, \bibinfo {author} {\bibfnamefont
  {P.}~\bibnamefont {Linse}}, \ and\ \bibinfo {author} {\bibfnamefont
  {P.}~\bibnamefont {Schurtenberger}},\ }\href@noop {} {\bibfield  {journal}
  {\bibinfo  {journal} {Nat. Commun.}\ }\textbf {\bibinfo {volume} {5}},\
  \bibinfo {pages} {5516} (\bibinfo {year} {2014})}\BibitemShut {NoStop}%
\bibitem [{\citenamefont {Sherman}\ \emph {et~al.}(2018)\citenamefont
  {Sherman}, \citenamefont {Ghosh},\ and\ \citenamefont
  {Swan}}]{Sherman.Z:2018}%
  \BibitemOpen
  \bibfield  {author} {\bibinfo {author} {\bibfnamefont {Z.~M.}\ \bibnamefont
  {Sherman}}, \bibinfo {author} {\bibfnamefont {D.}~\bibnamefont {Ghosh}}, \
  and\ \bibinfo {author} {\bibfnamefont {J.}~\bibnamefont {Swan}},\ }\href@noop
  {} {\bibfield  {journal} {\bibinfo  {journal} {Langmuir}\ }\textbf {\bibinfo
  {volume} {34}},\ \bibinfo {pages} {7117} (\bibinfo {year}
  {2018})}\BibitemShut {NoStop}%
\bibitem [{\citenamefont {Li}\ \emph {et~al.}(2006)\citenamefont {Li},
  \citenamefont {Yang}, \citenamefont {Bai},\ and\ \citenamefont
  {Huang}}]{Li.G:2006}%
  \BibitemOpen
  \bibfield  {author} {\bibinfo {author} {\bibfnamefont {G.}~\bibnamefont
  {Li}}, \bibinfo {author} {\bibfnamefont {X.}~\bibnamefont {Yang}}, \bibinfo
  {author} {\bibfnamefont {F.}~\bibnamefont {Bai}}, \ and\ \bibinfo {author}
  {\bibfnamefont {W.}~\bibnamefont {Huang}},\ }\href@noop {} {\bibfield
  {journal} {\bibinfo  {journal} {J. Colloid Interface Sci.}\ }\textbf
  {\bibinfo {volume} {297}},\ \bibinfo {pages} {705} (\bibinfo {year}
  {2006})}\BibitemShut {NoStop}%
\bibitem [{\citenamefont {Demir{\"o}rs}\ and\ \citenamefont
  {Alison}(2018)}]{Demirors.A:2018}%
  \BibitemOpen
  \bibfield  {author} {\bibinfo {author} {\bibfnamefont {A.~F.}\ \bibnamefont
  {Demir{\"o}rs}}\ and\ \bibinfo {author} {\bibfnamefont {L.}~\bibnamefont
  {Alison}},\ }\href@noop {} {\bibfield  {journal} {\bibinfo  {journal} {J.
  Phys. Chem. Lett.}\ }\textbf {\bibinfo {volume} {9}},\ \bibinfo {pages}
  {4437} (\bibinfo {year} {2018})}\BibitemShut {NoStop}%
\bibitem [{\citenamefont {Lin}\ \emph {et~al.}(2018)\citenamefont {Lin},
  \citenamefont {Hill}, \citenamefont {Peng},\ and\ \citenamefont
  {Zheng}}]{Lin.L:2018}%
  \BibitemOpen
  \bibfield  {author} {\bibinfo {author} {\bibfnamefont {L.}~\bibnamefont
  {Lin}}, \bibinfo {author} {\bibfnamefont {E.~H.}\ \bibnamefont {Hill}},
  \bibinfo {author} {\bibfnamefont {X.}~\bibnamefont {Peng}}, \ and\ \bibinfo
  {author} {\bibfnamefont {Y.}~\bibnamefont {Zheng}},\ }\href@noop {}
  {\bibfield  {journal} {\bibinfo  {journal} {Acc. Chem. Res.}\ }\textbf
  {\bibinfo {volume} {51}},\ \bibinfo {pages} {1465} (\bibinfo {year}
  {2018})}\BibitemShut {NoStop}%
\bibitem [{\citenamefont {Caciagli}\ \emph {et~al.}(2020)\citenamefont
  {Caciagli}, \citenamefont {Singh}, \citenamefont {Joshi}, \citenamefont
  {Adhikari},\ and\ \citenamefont {Eiser}}]{caciagli.a:2020}%
  \BibitemOpen
  \bibfield  {author} {\bibinfo {author} {\bibfnamefont {A.}~\bibnamefont
  {Caciagli}}, \bibinfo {author} {\bibfnamefont {R.}~\bibnamefont {Singh}},
  \bibinfo {author} {\bibfnamefont {D.}~\bibnamefont {Joshi}}, \bibinfo
  {author} {\bibfnamefont {R.}~\bibnamefont {Adhikari}}, \ and\ \bibinfo
  {author} {\bibfnamefont {E.}~\bibnamefont {Eiser}},\ }\href {\doibase
  10.1103/PhysRevLett.125.068001} {\bibfield  {journal} {\bibinfo  {journal}
  {Phys. Rev. Lett.}\ }\textbf {\bibinfo {volume} {125}},\ \bibinfo {pages}
  {068001} (\bibinfo {year} {2020})}\BibitemShut {NoStop}%
\bibitem [{\citenamefont {Pedreroa}\ and\ \citenamefont
  {Tierno}(2018)}]{Pedreroa.F:2018}%
  \BibitemOpen
  \bibfield  {author} {\bibinfo {author} {\bibfnamefont {F.~M.}\ \bibnamefont
  {Pedreroa}}\ and\ \bibinfo {author} {\bibfnamefont {P.}~\bibnamefont
  {Tierno}},\ }\href@noop {} {\bibfield  {journal} {\bibinfo  {journal} {J.
  Colloid Interface Sci.}\ }\textbf {\bibinfo {volume} {519}},\ \bibinfo
  {pages} {296} (\bibinfo {year} {2018})}\BibitemShut {NoStop}%
\bibitem [{\citenamefont {Chiou}\ \emph {et~al.}(2005)\citenamefont {Chiou},
  \citenamefont {Ohta},\ and\ \citenamefont {Wu}}]{Chiou.P:2005}%
  \BibitemOpen
  \bibfield  {author} {\bibinfo {author} {\bibfnamefont {P.~Y.}\ \bibnamefont
  {Chiou}}, \bibinfo {author} {\bibfnamefont {A.~T.}\ \bibnamefont {Ohta}}, \
  and\ \bibinfo {author} {\bibfnamefont {M.~C.}\ \bibnamefont {Wu}},\
  }\href@noop {} {\bibfield  {journal} {\bibinfo  {journal} {Nature}\ }\textbf
  {\bibinfo {volume} {436}},\ \bibinfo {pages} {370} (\bibinfo {year}
  {2005})}\BibitemShut {NoStop}%
\bibitem [{\citenamefont {Dienerowitz}\ \emph {et~al.}(2008)\citenamefont
  {Dienerowitz}, \citenamefont {Mazilu},\ and\ \citenamefont
  {Dholakia}}]{Dienerowitz.M:2008}%
  \BibitemOpen
  \bibfield  {author} {\bibinfo {author} {\bibfnamefont {M.}~\bibnamefont
  {Dienerowitz}}, \bibinfo {author} {\bibfnamefont {M.}~\bibnamefont {Mazilu}},
  \ and\ \bibinfo {author} {\bibfnamefont {K.}~\bibnamefont {Dholakia}},\
  }\href@noop {} {\bibfield  {journal} {\bibinfo  {journal} {J. Nanophotonics}\
  }\textbf {\bibinfo {volume} {2}},\ \bibinfo {pages} {021875} (\bibinfo {year}
  {2008})}\BibitemShut {NoStop}%
\bibitem [{\citenamefont {Yang}\ \emph {et~al.}(2009)\citenamefont {Yang},
  \citenamefont {Moore}, \citenamefont {Schmidt}, \citenamefont {Klug},
  \citenamefont {Lipson},\ and\ \citenamefont {Erickson}}]{Yang.A:2009}%
  \BibitemOpen
  \bibfield  {author} {\bibinfo {author} {\bibfnamefont {A.~H.~J.}\
  \bibnamefont {Yang}}, \bibinfo {author} {\bibfnamefont {S.~D.}\ \bibnamefont
  {Moore}}, \bibinfo {author} {\bibfnamefont {B.~S.}\ \bibnamefont {Schmidt}},
  \bibinfo {author} {\bibfnamefont {M.}~\bibnamefont {Klug}}, \bibinfo {author}
  {\bibfnamefont {M.}~\bibnamefont {Lipson}}, \ and\ \bibinfo {author}
  {\bibfnamefont {D.}~\bibnamefont {Erickson}},\ }\href@noop {} {\bibfield
  {journal} {\bibinfo  {journal} {Nature}\ }\textbf {\bibinfo {volume} {457}},\
  \bibinfo {pages} {71} (\bibinfo {year} {2009})}\BibitemShut {NoStop}%
\bibitem [{\citenamefont {Rodriguez}\ and\ \citenamefont
  {Markx}(2004)}]{Rodriguez.N:2004}%
  \BibitemOpen
  \bibfield  {author} {\bibinfo {author} {\bibfnamefont {N.~F.}\ \bibnamefont
  {Rodriguez}}\ and\ \bibinfo {author} {\bibfnamefont {G.}~\bibnamefont
  {Markx}},\ }\href@noop {} {\bibfield  {journal} {\bibinfo  {journal} {J.
  Physics D: Appl. Phys.}\ }\textbf {\bibinfo {volume} {37}},\ \bibinfo {pages}
  {353} (\bibinfo {year} {2004})}\BibitemShut {NoStop}%
\bibitem [{\citenamefont {Lee}\ \emph {et~al.}(2001)\citenamefont {Lee},
  \citenamefont {Lee},\ and\ \citenamefont {Westervelt}}]{Lee.C:2001}%
  \BibitemOpen
  \bibfield  {author} {\bibinfo {author} {\bibfnamefont {C.~S.}\ \bibnamefont
  {Lee}}, \bibinfo {author} {\bibfnamefont {H.}~\bibnamefont {Lee}}, \ and\
  \bibinfo {author} {\bibfnamefont {R.~M.}\ \bibnamefont {Westervelt}},\
  }\href@noop {} {\bibfield  {journal} {\bibinfo  {journal} {Appl. Phys.
  Lett.}\ }\textbf {\bibinfo {volume} {79}},\ \bibinfo {pages} {3308} (\bibinfo
  {year} {2001})}\BibitemShut {NoStop}%
\bibitem [{\citenamefont {Ruan}\ \emph {et~al.}(2010)\citenamefont {Ruan},
  \citenamefont {Vieira}, \citenamefont {Henighan}, \citenamefont {Chen},
  \citenamefont {Thakur}, \citenamefont {Sooryakumar},\ and\ \citenamefont
  {Winter}}]{Ruan.G:2010}%
  \BibitemOpen
  \bibfield  {author} {\bibinfo {author} {\bibfnamefont {G.}~\bibnamefont
  {Ruan}}, \bibinfo {author} {\bibfnamefont {G.}~\bibnamefont {Vieira}},
  \bibinfo {author} {\bibfnamefont {T.}~\bibnamefont {Henighan}}, \bibinfo
  {author} {\bibfnamefont {A.}~\bibnamefont {Chen}}, \bibinfo {author}
  {\bibfnamefont {D.}~\bibnamefont {Thakur}}, \bibinfo {author} {\bibfnamefont
  {R.}~\bibnamefont {Sooryakumar}}, \ and\ \bibinfo {author} {\bibfnamefont
  {J.~O.}\ \bibnamefont {Winter}},\ }\href@noop {} {\bibfield  {journal}
  {\bibinfo  {journal} {Nano Lett.}\ }\textbf {\bibinfo {volume} {10}},\
  \bibinfo {pages} {2220} (\bibinfo {year} {2010})}\BibitemShut {NoStop}%
\bibitem [{\citenamefont {Lv}\ \emph {et~al.}(2018)\citenamefont {Lv},
  \citenamefont {Varanakkottu}, \citenamefont {Baier},\ and\ \citenamefont
  {Hardt}}]{lv.c:2018}%
  \BibitemOpen
  \bibfield  {author} {\bibinfo {author} {\bibfnamefont {C.}~\bibnamefont
  {Lv}}, \bibinfo {author} {\bibfnamefont {C.~N.}\ \bibnamefont
  {Varanakkottu}}, \bibinfo {author} {\bibfnamefont {T.}~\bibnamefont {Baier}},
  \ and\ \bibinfo {author} {\bibfnamefont {S.}~\bibnamefont {Hardt}},\
  }\href@noop {} {\bibfield  {journal} {\bibinfo  {journal} {Nano Lett.}\
  }\textbf {\bibinfo {volume} {18}},\ \bibinfo {pages} {6924} (\bibinfo {year}
  {2018})}\BibitemShut {NoStop}%
\bibitem [{\citenamefont {Dom\'{\i}nguez}\ \emph {et~al.}(2016)\citenamefont
  {Dom\'{\i}nguez}, \citenamefont {Malgaretti}, \citenamefont {Popescu},\ and\
  \citenamefont {Dietrich}}]{dominiges.a:2016}%
  \BibitemOpen
  \bibfield  {author} {\bibinfo {author} {\bibfnamefont {A.}~\bibnamefont
  {Dom\'{\i}nguez}}, \bibinfo {author} {\bibfnamefont {P.}~\bibnamefont
  {Malgaretti}}, \bibinfo {author} {\bibfnamefont {M.~N.}\ \bibnamefont
  {Popescu}}, \ and\ \bibinfo {author} {\bibfnamefont {S.}~\bibnamefont
  {Dietrich}},\ }\href {\doibase 10.1103/PhysRevLett.116.078301} {\bibfield
  {journal} {\bibinfo  {journal} {Phys. Rev. Lett.}\ }\textbf {\bibinfo
  {volume} {116}},\ \bibinfo {pages} {078301} (\bibinfo {year}
  {2016})}\BibitemShut {NoStop}%
\bibitem [{\citenamefont {Peter}\ \emph {et~al.}(2020)\citenamefont {Peter},
  \citenamefont {Malgaretti}, \citenamefont {Rivas}, \citenamefont
  {Scagliarini}, \citenamefont {Harting},\ and\ \citenamefont
  {Dietrich}}]{peter.t:2020}%
  \BibitemOpen
  \bibfield  {author} {\bibinfo {author} {\bibfnamefont {T.}~\bibnamefont
  {Peter}}, \bibinfo {author} {\bibfnamefont {P.}~\bibnamefont {Malgaretti}},
  \bibinfo {author} {\bibfnamefont {N.}~\bibnamefont {Rivas}}, \bibinfo
  {author} {\bibfnamefont {A.}~\bibnamefont {Scagliarini}}, \bibinfo {author}
  {\bibfnamefont {J.}~\bibnamefont {Harting}}, \ and\ \bibinfo {author}
  {\bibfnamefont {S.}~\bibnamefont {Dietrich}},\ }\href {\doibase
  10.1039/C9SM02247C} {\bibfield  {journal} {\bibinfo  {journal} {Soft Matter}\
  }\textbf {\bibinfo {volume} {16}},\ \bibinfo {pages} {3536} (\bibinfo {year}
  {2020})}\BibitemShut {NoStop}%
\bibitem [{\citenamefont {Niu}\ \emph {et~al.}(2017{\natexlab{a}})\citenamefont
  {Niu}, \citenamefont {Palberg},\ and\ \citenamefont {Speck}}]{niu.r:2017}%
  \BibitemOpen
  \bibfield  {author} {\bibinfo {author} {\bibfnamefont {R.}~\bibnamefont
  {Niu}}, \bibinfo {author} {\bibfnamefont {T.}~\bibnamefont {Palberg}}, \ and\
  \bibinfo {author} {\bibfnamefont {T.}~\bibnamefont {Speck}},\ }\href
  {\doibase 10.1103/PhysRevLett.119.028001} {\bibfield  {journal} {\bibinfo
  {journal} {Phys. Rev. Lett.}\ }\textbf {\bibinfo {volume} {119}},\ \bibinfo
  {pages} {028001} (\bibinfo {year} {2017}{\natexlab{a}})}\BibitemShut
  {NoStop}%
\bibitem [{\citenamefont {Niu}\ \emph {et~al.}(2017{\natexlab{b}})\citenamefont
  {Niu}, \citenamefont {Oguz}, \citenamefont {Müller}, \citenamefont
  {Reinmüller}, \citenamefont {Botin}, \citenamefont {Löwen},\ and\
  \citenamefont {Palberg}}]{niu.r:2017b}%
  \BibitemOpen
  \bibfield  {author} {\bibinfo {author} {\bibfnamefont {R.}~\bibnamefont
  {Niu}}, \bibinfo {author} {\bibfnamefont {E.~C.}\ \bibnamefont {Oguz}},
  \bibinfo {author} {\bibfnamefont {H.}~\bibnamefont {Müller}}, \bibinfo
  {author} {\bibfnamefont {A.}~\bibnamefont {Reinmüller}}, \bibinfo {author}
  {\bibfnamefont {D.}~\bibnamefont {Botin}}, \bibinfo {author} {\bibfnamefont
  {H.}~\bibnamefont {Löwen}}, \ and\ \bibinfo {author} {\bibfnamefont
  {T.}~\bibnamefont {Palberg}},\ }\href {\doibase 10.1039/C6CP07231C}
  {\bibfield  {journal} {\bibinfo  {journal} {Phys. Chem. Chem. Phys.}\
  }\textbf {\bibinfo {volume} {19}},\ \bibinfo {pages} {3104} (\bibinfo {year}
  {2017}{\natexlab{b}})}\BibitemShut {NoStop}%
\bibitem [{\citenamefont {Prieve}\ \emph {et~al.}(1984)\citenamefont {Prieve},
  \citenamefont {Anderson}, \citenamefont {Ebel},\ and\ \citenamefont
  {Lowell}}]{prieve.dc:1984}%
  \BibitemOpen
  \bibfield  {author} {\bibinfo {author} {\bibfnamefont {D.~C.}\ \bibnamefont
  {Prieve}}, \bibinfo {author} {\bibfnamefont {J.~L.}\ \bibnamefont
  {Anderson}}, \bibinfo {author} {\bibfnamefont {J.~P.}\ \bibnamefont {Ebel}},
  \ and\ \bibinfo {author} {\bibfnamefont {M.~E.}\ \bibnamefont {Lowell}},\
  }\href@noop {} {\bibfield  {journal} {\bibinfo  {journal} {J. Fluid Mech.}\
  }\textbf {\bibinfo {volume} {148}},\ \bibinfo {pages} {247} (\bibinfo {year}
  {1984})}\BibitemShut {NoStop}%
\bibitem [{\citenamefont {Marbach}\ and\ \citenamefont
  {Bocquet}(2019)}]{Marbach.S:2017}%
  \BibitemOpen
  \bibfield  {author} {\bibinfo {author} {\bibfnamefont {S.}~\bibnamefont
  {Marbach}}\ and\ \bibinfo {author} {\bibfnamefont {L.}~\bibnamefont
  {Bocquet}},\ }\href@noop {} {\bibfield  {journal} {\bibinfo  {journal} {Chem.
  Soc. Rev.}\ }\textbf {\bibinfo {volume} {48}},\ \bibinfo {pages} {3102}
  (\bibinfo {year} {2019})}\BibitemShut {NoStop}%
\bibitem [{\citenamefont {McDermott}\ \emph {et~al.}(2012)\citenamefont
  {McDermott}, \citenamefont {Kar}, \citenamefont {Daher}, \citenamefont
  {Klara}, \citenamefont {Wang}, \citenamefont {Sen},\ and\ \citenamefont
  {Velegol}}]{mcdermott2012}%
  \BibitemOpen
  \bibfield  {author} {\bibinfo {author} {\bibfnamefont {J.~J.}\ \bibnamefont
  {McDermott}}, \bibinfo {author} {\bibfnamefont {A.}~\bibnamefont {Kar}},
  \bibinfo {author} {\bibfnamefont {M.}~\bibnamefont {Daher}}, \bibinfo
  {author} {\bibfnamefont {S.}~\bibnamefont {Klara}}, \bibinfo {author}
  {\bibfnamefont {G.}~\bibnamefont {Wang}}, \bibinfo {author} {\bibfnamefont
  {A.}~\bibnamefont {Sen}}, \ and\ \bibinfo {author} {\bibfnamefont
  {D.}~\bibnamefont {Velegol}},\ }\href@noop {} {\bibfield  {journal} {\bibinfo
   {journal} {Langmuir}\ }\textbf {\bibinfo {volume} {28}},\ \bibinfo {pages}
  {15491} (\bibinfo {year} {2012})}\BibitemShut {NoStop}%
\bibitem [{\citenamefont {Feldmann}\ \emph {et~al.}(2016)\citenamefont
  {Feldmann}, \citenamefont {Maduar}, \citenamefont {Santer}, \citenamefont
  {Lomadze}, \citenamefont {Vinogradova},\ and\ \citenamefont
  {Santer}}]{feldmann2016}%
  \BibitemOpen
  \bibfield  {author} {\bibinfo {author} {\bibfnamefont {D.}~\bibnamefont
  {Feldmann}}, \bibinfo {author} {\bibfnamefont {S.~R.}\ \bibnamefont
  {Maduar}}, \bibinfo {author} {\bibfnamefont {M.}~\bibnamefont {Santer}},
  \bibinfo {author} {\bibfnamefont {N.}~\bibnamefont {Lomadze}}, \bibinfo
  {author} {\bibfnamefont {O.~I.}\ \bibnamefont {Vinogradova}}, \ and\ \bibinfo
  {author} {\bibfnamefont {S.}~\bibnamefont {Santer}},\ }\href@noop {}
  {\bibfield  {journal} {\bibinfo  {journal} {Sci. Rep.}\ }\textbf {\bibinfo
  {volume} {6}},\ \bibinfo {pages} {1} (\bibinfo {year} {2016})}\BibitemShut
  {NoStop}%
\bibitem [{\citenamefont {Feldmann}\ \emph {et~al.}(2020)\citenamefont
  {Feldmann}, \citenamefont {Arya}, \citenamefont {Molotilin}, \citenamefont
  {Lomadze}, \citenamefont {Kopyshev}, \citenamefont {Vinogradova},\ and\
  \citenamefont {Santer}}]{feldmann.d:2020}%
  \BibitemOpen
  \bibfield  {author} {\bibinfo {author} {\bibfnamefont {D.}~\bibnamefont
  {Feldmann}}, \bibinfo {author} {\bibfnamefont {P.}~\bibnamefont {Arya}},
  \bibinfo {author} {\bibfnamefont {T.~Y.}\ \bibnamefont {Molotilin}}, \bibinfo
  {author} {\bibfnamefont {N.}~\bibnamefont {Lomadze}}, \bibinfo {author}
  {\bibfnamefont {A.}~\bibnamefont {Kopyshev}}, \bibinfo {author}
  {\bibfnamefont {O.~I.}\ \bibnamefont {Vinogradova}}, \ and\ \bibinfo {author}
  {\bibfnamefont {S.~A.}\ \bibnamefont {Santer}},\ }\href@noop {} {\bibfield
  {journal} {\bibinfo  {journal} {Langmuir}\ }\textbf {\bibinfo {volume}
  {36}},\ \bibinfo {pages} {6994} (\bibinfo {year} {2020})}\BibitemShut
  {NoStop}%
\bibitem [{\citenamefont {Williams}\ \emph {et~al.}(1992)\citenamefont
  {Williams}, \citenamefont {Koch},\ and\ \citenamefont
  {Giddings}}]{Williams1992}%
  \BibitemOpen
  \bibfield  {author} {\bibinfo {author} {\bibfnamefont {P.~S.}\ \bibnamefont
  {Williams}}, \bibinfo {author} {\bibfnamefont {T.}~\bibnamefont {Koch}}, \
  and\ \bibinfo {author} {\bibfnamefont {J.~C.}\ \bibnamefont {Giddings}},\
  }\href@noop {} {\bibfield  {journal} {\bibinfo  {journal} {Chem. Eng.
  Commun.}\ }\textbf {\bibinfo {volume} {111}},\ \bibinfo {pages} {121}
  (\bibinfo {year} {1992})}\BibitemShut {NoStop}%
\bibitem [{\citenamefont {Asmolov}\ \emph {et~al.}(2015)\citenamefont
  {Asmolov}, \citenamefont {Dubov}, \citenamefont {Nizkaya}, \citenamefont
  {Kuehne},\ and\ \citenamefont {Vinogradova}}]{asmolov2015}%
  \BibitemOpen
  \bibfield  {author} {\bibinfo {author} {\bibfnamefont {E.~S.}\ \bibnamefont
  {Asmolov}}, \bibinfo {author} {\bibfnamefont {A.~L.}\ \bibnamefont {Dubov}},
  \bibinfo {author} {\bibfnamefont {T.~V.}\ \bibnamefont {Nizkaya}}, \bibinfo
  {author} {\bibfnamefont {A.~J.~C.}\ \bibnamefont {Kuehne}}, \ and\ \bibinfo
  {author} {\bibfnamefont {O.~I.}\ \bibnamefont {Vinogradova}},\ }\href@noop {}
  {\bibfield  {journal} {\bibinfo  {journal} {Lab. Chip}\ }\textbf {\bibinfo
  {volume} {15}},\ \bibinfo {pages} {2835} (\bibinfo {year}
  {2015})}\BibitemShut {NoStop}%
\bibitem [{\citenamefont {Dubov}\ \emph {et~al.}(2017)\citenamefont {Dubov},
  \citenamefont {Molotilin},\ and\ \citenamefont
  {Vinogradova}}]{dubov.al:2017}%
  \BibitemOpen
  \bibfield  {author} {\bibinfo {author} {\bibfnamefont {A.~L.}\ \bibnamefont
  {Dubov}}, \bibinfo {author} {\bibfnamefont {T.~Y.}\ \bibnamefont
  {Molotilin}}, \ and\ \bibinfo {author} {\bibfnamefont {O.~I.}\ \bibnamefont
  {Vinogradova}},\ }\href {\doibase 10.1039/C7SM00986K} {\bibfield  {journal}
  {\bibinfo  {journal} {Soft Matter}\ }\textbf {\bibinfo {volume} {13}},\
  \bibinfo {pages} {7498} (\bibinfo {year} {2017})}\BibitemShut {NoStop}%
\bibitem [{\citenamefont {{Vinogradova}}\ \emph {et~al.}(2020)\citenamefont
  {{Vinogradova}}, \citenamefont {{Silkina}}, \citenamefont {Bag},\ and\
  \citenamefont {Asmolov}}]{vinogradova.oi:2020}%
  \BibitemOpen
  \bibfield  {author} {\bibinfo {author} {\bibfnamefont {O.~I.}\ \bibnamefont
  {{Vinogradova}}}, \bibinfo {author} {\bibfnamefont {E.~F.}\ \bibnamefont
  {{Silkina}}}, \bibinfo {author} {\bibfnamefont {N.}~\bibnamefont {Bag}}, \
  and\ \bibinfo {author} {\bibfnamefont {E.~S.}\ \bibnamefont {Asmolov}},\
  }\href@noop {} {\bibfield  {journal} {\bibinfo  {journal} {Phys. Fluids}\
  }\textbf {\bibinfo {volume} {32}},\ \bibinfo {pages} {102105} (\bibinfo
  {year} {2020})}\BibitemShut {NoStop}%
\bibitem [{\citenamefont {Silkina}\ \emph {et~al.}(2020)\citenamefont
  {Silkina}, \citenamefont {Molotilin}, \citenamefont {Maduar},\ and\
  \citenamefont {Vinogradova}}]{silkina.ef:2020}%
  \BibitemOpen
  \bibfield  {author} {\bibinfo {author} {\bibfnamefont {E.~F.}\ \bibnamefont
  {Silkina}}, \bibinfo {author} {\bibfnamefont {T.~Y.}\ \bibnamefont
  {Molotilin}}, \bibinfo {author} {\bibfnamefont {S.~R.}\ \bibnamefont
  {Maduar}}, \ and\ \bibinfo {author} {\bibfnamefont {O.~I.}\ \bibnamefont
  {Vinogradova}},\ }\href@noop {} {\bibfield  {journal} {\bibinfo  {journal}
  {Soft Matter}\ }\textbf {\bibinfo {volume} {16}},\ \bibinfo {pages} {929}
  (\bibinfo {year} {2020})}\BibitemShut {NoStop}%
\bibitem [{\citenamefont {Rumyantsev}\ \emph {et~al.}(2014)\citenamefont
  {Rumyantsev}, \citenamefont {Santer},\ and\ \citenamefont
  {Kramarenko}}]{rumyantsev.a:2014}%
  \BibitemOpen
  \bibfield  {author} {\bibinfo {author} {\bibfnamefont {A.~M.}\ \bibnamefont
  {Rumyantsev}}, \bibinfo {author} {\bibfnamefont {S.}~\bibnamefont {Santer}},
  \ and\ \bibinfo {author} {\bibfnamefont {E.~Y.}\ \bibnamefont {Kramarenko}},\
  }\href@noop {} {\bibfield  {journal} {\bibinfo  {journal} {Macromolecules}\
  }\textbf {\bibinfo {volume} {47}},\ \bibinfo {pages} {5388} (\bibinfo {year}
  {2014})}\BibitemShut {NoStop}%
\bibitem [{\citenamefont {Schimka}\ \emph
  {et~al.}(2017{\natexlab{a}})\citenamefont {Schimka}, \citenamefont {Lomadze},
  \citenamefont {Rabe}, \citenamefont {Kopyshev}, \citenamefont {Lehmann},
  \citenamefont {Klitzing}, \citenamefont {Rumyantsev}, \citenamefont
  {Kramarenko},\ and\ \citenamefont {Santer}}]{schimka.s:2017a}%
  \BibitemOpen
  \bibfield  {author} {\bibinfo {author} {\bibfnamefont {S.}~\bibnamefont
  {Schimka}}, \bibinfo {author} {\bibfnamefont {N.}~\bibnamefont {Lomadze}},
  \bibinfo {author} {\bibfnamefont {M.}~\bibnamefont {Rabe}}, \bibinfo {author}
  {\bibfnamefont {A.}~\bibnamefont {Kopyshev}}, \bibinfo {author}
  {\bibfnamefont {M.}~\bibnamefont {Lehmann}}, \bibinfo {author} {\bibfnamefont
  {R.~V.}\ \bibnamefont {Klitzing}}, \bibinfo {author} {\bibfnamefont {A.~M.}\
  \bibnamefont {Rumyantsev}}, \bibinfo {author} {\bibfnamefont {E.~Y.}\
  \bibnamefont {Kramarenko}}, \ and\ \bibinfo {author} {\bibfnamefont
  {S.}~\bibnamefont {Santer}},\ }\href@noop {} {\bibfield  {journal} {\bibinfo
  {journal} {Phys. Chem. Chem. Phys.}\ }\textbf {\bibinfo {volume} {19}},\
  \bibinfo {pages} {108} (\bibinfo {year} {2017}{\natexlab{a}})}\BibitemShut
  {NoStop}%
\bibitem [{\citenamefont {Schimka}\ \emph
  {et~al.}(2017{\natexlab{b}})\citenamefont {Schimka}, \citenamefont
  {Gordievskaya}, \citenamefont {Lomadze}, \citenamefont {Lehmann},
  \citenamefont {Klitzing}, \citenamefont {Rumyantsev}, \citenamefont
  {Kramarenko},\ and\ \citenamefont {Santer}}]{schimka.s:2017b}%
  \BibitemOpen
  \bibfield  {author} {\bibinfo {author} {\bibfnamefont {S.}~\bibnamefont
  {Schimka}}, \bibinfo {author} {\bibfnamefont {Y.~D.}\ \bibnamefont
  {Gordievskaya}}, \bibinfo {author} {\bibfnamefont {N.}~\bibnamefont
  {Lomadze}}, \bibinfo {author} {\bibfnamefont {M.}~\bibnamefont {Lehmann}},
  \bibinfo {author} {\bibfnamefont {R.~V.}\ \bibnamefont {Klitzing}}, \bibinfo
  {author} {\bibfnamefont {A.~M.}\ \bibnamefont {Rumyantsev}}, \bibinfo
  {author} {\bibfnamefont {E.~Y.}\ \bibnamefont {Kramarenko}}, \ and\ \bibinfo
  {author} {\bibfnamefont {S.}~\bibnamefont {Santer}},\ }\href@noop {}
  {\bibfield  {journal} {\bibinfo  {journal} {J. Chem. Phys.}\ }\textbf
  {\bibinfo {volume} {147}},\ \bibinfo {pages} {031101} (\bibinfo {year}
  {2017}{\natexlab{b}})}\BibitemShut {NoStop}%
\bibitem [{\citenamefont {Beavers}\ and\ \citenamefont
  {Joseph}(1967)}]{beavers.gs:1967}%
  \BibitemOpen
  \bibfield  {author} {\bibinfo {author} {\bibfnamefont {G.~S.}\ \bibnamefont
  {Beavers}}\ and\ \bibinfo {author} {\bibfnamefont {D.~D.}\ \bibnamefont
  {Joseph}},\ }\href@noop {} {\bibfield  {journal} {\bibinfo  {journal} {J.
  Fluid Mech.}\ }\textbf {\bibinfo {volume} {30}},\ \bibinfo {pages} {197}
  (\bibinfo {year} {1967})}\BibitemShut {NoStop}%
\bibitem [{\citenamefont {Feldmann}\ \emph {et~al.}(2019)\citenamefont
  {Feldmann}, \citenamefont {Arya}, \citenamefont {Lomadze}, \citenamefont
  {Kopyshev},\ and\ \citenamefont {Santer}}]{Feldmann.D:2019}%
  \BibitemOpen
  \bibfield  {author} {\bibinfo {author} {\bibfnamefont {D.}~\bibnamefont
  {Feldmann}}, \bibinfo {author} {\bibfnamefont {P.}~\bibnamefont {Arya}},
  \bibinfo {author} {\bibfnamefont {N.}~\bibnamefont {Lomadze}}, \bibinfo
  {author} {\bibfnamefont {A.}~\bibnamefont {Kopyshev}}, \ and\ \bibinfo
  {author} {\bibfnamefont {S.}~\bibnamefont {Santer}},\ }\href@noop {}
  {\bibfield  {journal} {\bibinfo  {journal} {App. Phys. Lett.}\ }\textbf
  {\bibinfo {volume} {115}},\ \bibinfo {pages} {263701} (\bibinfo {year}
  {2019})}\BibitemShut {NoStop}%
\bibitem [{\citenamefont {Arya}\ \emph
  {et~al.}(2020{\natexlab{a}})\citenamefont {Arya}, \citenamefont {Jelken},
  \citenamefont {Feldmann}, \citenamefont {Lomadze},\ and\ \citenamefont
  {Santer}}]{Arya.P:2020b}%
  \BibitemOpen
  \bibfield  {author} {\bibinfo {author} {\bibfnamefont {P.}~\bibnamefont
  {Arya}}, \bibinfo {author} {\bibfnamefont {J.}~\bibnamefont {Jelken}},
  \bibinfo {author} {\bibfnamefont {D.}~\bibnamefont {Feldmann}}, \bibinfo
  {author} {\bibfnamefont {N.}~\bibnamefont {Lomadze}}, \ and\ \bibinfo
  {author} {\bibfnamefont {S.}~\bibnamefont {Santer}},\ }\href@noop {}
  {\bibfield  {journal} {\bibinfo  {journal} {J. Chem. Phys.}\
  }\textbf {\bibinfo {volume} {152}},\ \bibinfo {pages} {194703} (\bibinfo
  {year} {2020}{\natexlab{a}})}\BibitemShut {NoStop}%
\bibitem [{\citenamefont {Dumont}\ \emph {et~al.}(2002)\citenamefont {Dumont},
  \citenamefont {Galstian}, \citenamefont {Senkow},\ and\ \citenamefont
  {Ritcey}}]{Dumont.D:2002}%
  \BibitemOpen
  \bibfield  {author} {\bibinfo {author} {\bibfnamefont {D.}~\bibnamefont
  {Dumont}}, \bibinfo {author} {\bibfnamefont {T.~V.}\ \bibnamefont
  {Galstian}}, \bibinfo {author} {\bibfnamefont {S.}~\bibnamefont {Senkow}}, \
  and\ \bibinfo {author} {\bibfnamefont {A.~M.}\ \bibnamefont {Ritcey}},\
  }\href@noop {} {\bibfield  {journal} {\bibinfo  {journal} {Mol. Cryst. Liq.
  Cryst.}\ }\textbf {\bibinfo {volume} {375}},\ \bibinfo {pages} {341}
  (\bibinfo {year} {2002})}\BibitemShut {NoStop}%
\bibitem [{\citenamefont {Arya}\ \emph
  {et~al.}(2020{\natexlab{b}})\citenamefont {Arya}, \citenamefont {Jelken},
  \citenamefont {Lomadze}, \citenamefont {Santer},\ and\ \citenamefont
  {Bekir}}]{Arya.P:2020a}%
  \BibitemOpen
  \bibfield  {author} {\bibinfo {author} {\bibfnamefont {P.}~\bibnamefont
  {Arya}}, \bibinfo {author} {\bibfnamefont {J.}~\bibnamefont {Jelken}},
  \bibinfo {author} {\bibfnamefont {N.}~\bibnamefont {Lomadze}}, \bibinfo
  {author} {\bibfnamefont {S.}~\bibnamefont {Santer}}, \ and\ \bibinfo {author}
  {\bibfnamefont {M.}~\bibnamefont {Bekir}},\ }\href@noop {} {\bibfield
  {journal} {\bibinfo  {journal} {J. Chem. Phys.}\ }\textbf {\bibinfo {volume}
  {152}},\ \bibinfo {pages} {024904} (\bibinfo {year}
  {2020}{\natexlab{b}})}\BibitemShut {NoStop}%
\bibitem [{\citenamefont {Sbalzarini}\ and\ \citenamefont
  {Koumoutsakos}(2005)}]{Sbalzarini.I:2005}%
  \BibitemOpen
  \bibfield  {author} {\bibinfo {author} {\bibfnamefont {I.~F.}\ \bibnamefont
  {Sbalzarini}}\ and\ \bibinfo {author} {\bibfnamefont {P.}~\bibnamefont
  {Koumoutsakos}},\ }\href@noop {} {\bibfield  {journal} {\bibinfo  {journal}
  {J. Struct. Biol.}\ }\textbf {\bibinfo {volume} {151}},\ \bibinfo {pages}
  {182} (\bibinfo {year} {2005})}\BibitemShut {NoStop}%
\bibitem [{\citenamefont {Arya}\ \emph
  {et~al.}(2020{\natexlab{c}})\citenamefont {Arya}, \citenamefont {Feldmann},
  \citenamefont {Kopyshev}, \citenamefont {Lomadze},\ and\ \citenamefont
  {Santer}}]{Arya.P:2020c}%
  \BibitemOpen
  \bibfield  {author} {\bibinfo {author} {\bibfnamefont {P.}~\bibnamefont
  {Arya}}, \bibinfo {author} {\bibfnamefont {D.}~\bibnamefont {Feldmann}},
  \bibinfo {author} {\bibfnamefont {A.}~\bibnamefont {Kopyshev}}, \bibinfo
  {author} {\bibfnamefont {N.}~\bibnamefont {Lomadze}}, \ and\ \bibinfo
  {author} {\bibfnamefont {S.}~\bibnamefont {Santer}},\ }\href@noop {}
  {\bibfield  {journal} {\bibinfo  {journal} {Soft Matter}\ }\textbf {\bibinfo
  {volume} {16}},\ \bibinfo {pages} {1148} (\bibinfo {year}
  {2020}{\natexlab{c}})}\BibitemShut {NoStop}%
\end{thebibliography}

%

\end{document}